\documentclass{article}

\usepackage[english]{babel}

\usepackage[letterpaper,top=2cm,bottom=2cm,left=3cm,right=3cm,marginparwidth=1.75cm]{geometry}

\usepackage{amsmath}
\usepackage{graphicx}
\usepackage[colorlinks=true, allcolors=blue]{hyperref}
\usepackage{bm}
\usepackage{float}
\usepackage{hyperref}
\usepackage{bbold}
\usepackage{verbatim}
\usepackage[table]{xcolor}
\definecolor{highlight}{RGB}{135,206,250} 
\usepackage{comment}
\usepackage{subfigure}

\usepackage{multirow}
\usepackage{booktabs}
\usepackage{natbib}
\usepackage{tikz}                 
\usetikzlibrary{shapes.geometric, arrows, shadows, fadings}
\tikzstyle{startstop} = [rectangle, rounded corners, minimum width=3.5cm, minimum height=1.2cm, text centered, draw=black, fill=gray!20]
\tikzstyle{arrow} = [thick,->,>=stealth]

\title{Bayesian blockwise inference for joint models of longitudinal and multistate processes}
\author{Sida Chen, Danilo Alvares, Christopher Jackson, Jessica Barrett\\
MRC Biostatistics Unit, University of Cambridge}

\begin{document}
\maketitle

\begin{abstract}
Joint models (JM) for longitudinal and survival data have gained increasing interest and found applications in a wide range of clinical and biomedical settings. These models facilitate the understanding of the relationship between outcomes and enable individualized predictions. In many applications, more complex event processes arise, necessitating joint longitudinal and multistate models. However, their practical application can be hindered by computational challenges due to increased model complexity and large sample sizes. Motivated by a longitudinal multimorbidity analysis of large UK health records, we have developed a scalable Bayesian methodology for such joint multistate models that is capable of handling complex event processes and large datasets, with straightforward implementation. We propose two blockwise inference approaches for different inferential purposes based on different levels of decomposition of the multistate processes. These approaches leverage parallel computing, ease the specification of different models for different transitions, and model/variable selection can be performed within a Bayesian framework using Bayesian leave-one-out cross-validation. Using a simulation study, we show that the proposed approaches achieve satisfactory performance regarding posterior point and interval estimation, with notable gains in sampling efficiency compared to the standard estimation strategy. We illustrate our approaches using a large UK electronic health record dataset where we analysed the coevolution of routinely measured systolic blood pressure (SBP) and the progression of multimorbidity, defined as the combinations of three chronic conditions. Our analysis identified distinct association structures between SBP and different disease transitions. 
\end{abstract}

\section{Introduction}

Joint models for longitudinal and time-to-event data have gained increasing interest in clinical and biomedical research \citep{rizopoulos2012joint}. 
These models are particularly relevant when subjects are followed over a period of time, during which repeated measurements of some markers are collected. The key objective is often to analyse the association between the marker and the risk of an event of interest, e.g. death. In the joint modelling framework, both the longitudinal marker and the event are treated as outcomes, with separate regression models employed for each process. Statistical inference is based on the joint distribution of both outcome processes, which is typically formulated via sharing parameters between the submodels for each process. Joint models effectively handle the endogenous, error-prone nature of the marker, as well as non-random dropout due to the occurrence of an event. Theoretical and empirical evidence show that joint modelling could achieve more accurate and efficient inference for each outcome process compared to analysing each process separately \citep{verbeke2009joint,ibrahim2010basic}.

The basic formulation of the model involves a single time-to-event outcome. However, in many application contexts, more complicated event processes may arise. For instance, a subject of interest may experience a succession of intermediate events, each representing a disease/health status. Multistate models (MSM) extend the traditional survival model to encompass more states, providing a useful probabilistic framework for modelling complex longitudinal event history data \citep{cook2018multistate}. Well-known models such as competing risks and illness-death models are special cases of MSM. MSM can be conveniently specified via a collection of time-to-event models, one for each permitted transition between two states. They have recently been incorporated into joint models, typically by linking the longitudinal and transition-specific regression models via shared random effects, allowing inference of association patterns between markers and different transitions. Examples of applications of such joint multistate models can be found in e.g. \citet{dantan2011joint} for jointly studying cognitive decline, risks of dementia and death, \citet{ferrer2016joint} for analysing the progression of prostate cancer, and \citet{dessie2020modelling} for predicting the clinical progression of HIV infection. See also \citet{hickey2018comparison} and references therein for works on joint competing risks models. Inference for such models has conventionally been based on maximum likelihood estimation theory. Alternatively, a Bayesian framework, which allows for the incorporation of prior knowledge and provides more coherent uncertainty quantification and dynamic prediction, has been explored more recently in \citet{furgal2021bayesian}. However, existing estimation methods under both frameworks face significant computational challenges with large sample sizes, and when the submodels become complex, for instance, in the presence of multivariate longitudinal markers and/or when the number of states/transitions in the MSM becomes large \citep{hickey2016joint,hickey2018joint}. For the frequentist approach, the main difficulty with the inference arises from the numerical approximation of the integral over the random effects, which is required in computing the likelihood function. In the Bayesian approach, posterior inference is typically based on Markov chain Monte Carlo (MCMC) sampling methods, which can be computationally expensive and sometimes face difficulties with mixing and convergence. Therefore, the application of such joint multistate models is currently limited to relatively small sample sizes and/or simple longitudinal and multistate structures due to computational limitations.

The aim of this paper is to propose new methodologies for scalable Bayesian inference in joint longitudinal and multistate models that can handle complex event processes and large data sets, with straightforward implementation. Instead of working with a joint likelihood function constructed based on the entire longitudinal and multistate data, as the standard approach would do, we propose two blockwise approaches. These approaches decompose the original estimation task into smaller inference blocks based on the multistate transition pattern, with parameters associated with each block estimated in a parallel manner, based solely on the longitudinal and time-to-event data associated with that block. More specifically, the first approach employs competing risk decompositions of the multistate process, estimating a joint longitudinal and competing risk model for each block. When focusing specifically on the association and other multistate parameters, our second block inference strategy offers further efficiency gains by utilizing transition-specific posteriors. Blockwise approaches also facilitate the specification of different models for different types of transitions (e.g. the association structure between the longitudinal and event processes). Model selection can be conveniently and efficiently performed in a block-specific manner using Bayesian leave-one-out cross-validation. It should be noted that these parallel inference strategies limit the ability to share information across different blocks, but this is more relevant when data are scarce. Through a simulation study, we compare the proposed approaches with the standard estimation strategy, all of which were implemented using the state-of-the-art sampling technique. Our methodologies demonstrate satisfactory performance in terms of posterior point and interval estimation, along with notable gains in computational efficiency. Additionally, the proposed approaches inherently improve robustness against longitudinal model misspecification due to the use of ``local'' rather than ``global'' longitudinal data when there are structural changes in the marker dynamics. We illustrate our approaches using a large anonymised dataset of electronic primary care records in England, the Clinical Practice Research Datalink (CPRD) Aurum \citep{wolf2019data}, where we analysed the coevolution of routinely measured systolic blood pressure (SBP) and the progression of multimorbidity, defined as the combinations of three chronic conditions. While multistate models have been effectively used in longitudinal studies of multimorbidity \citep{singh2018clinical,freisling2020lifestyle}, to our knowledge, no previous work has examined the association between longitudinal biomarkers and multimorbidity progression within a joint modelling framework. Our proposed approaches enable us to utilize a much larger volume of subjects' data during inference compared to the standard approach, facilitating the identification of differing association patterns between SBP and different disease transitions.

The manuscript is structured as follows. Section \ref{sec:jm-msm} introduces the basic Bayesian joint multistate models and the corresponding estimation method. In Section \ref{sec:block}, we present the proposed blockwise inference approaches, and their performance is evaluated through a simulation study in Section \ref{sec:sim}. In Section \ref{sec:appl}, we illustrate the application of the proposed approaches in a longitudinal multimorbidity analysis of large UK health records. Finally, Section \ref{sec:disc} provides a discussion and possible directions for further work.

\section{The Bayesian joint multistate model}
\label{sec:jm-msm}
We consider a joint model for a longitudinal process measured over time and a continuous-time multistate process. The two processes are modelled using distinct regression submodels which are linked by sharing subject-specific parameters. 

\subsection{Longitudinal submodel}

Let $y_{i}(t)$ be the value of the longitudinal process of subject $i$ measured at time $t$, and let $y_{i}=(y_{i1},\ldots,y_{in_{i}})$ be the observed $n_{i}$-dimensional longitudinal response vector for the subject, where $y_{ij} = y_{i}(t_{ij})$, $j=1,\ldots,n_{i}$.
We assume a mixed-effect model for the longitudinal process:
\begin{equation}
\begin{aligned}
  y_{i}(t) \mid {\bm b}_{i}, {\bm \beta}, \sigma^{2}  & \sim \text{Normal}\big(\mu_{i}(t \mid {\bm b}_{i}, {\bm \beta}), \sigma^{2}\big),\\
 {\bm b}_{i} \mid {\bm \Sigma}_{r \times r} &  \sim \text{Normal}({\bm 0}, {\bm\Sigma}_{r \times r}), \label{eq:REM}
\end{aligned}
\end{equation} 

\noindent where $\mu_{i}(t \mid {\bm b}_{i}, {\bm \beta})$ denotes a subject-specific (unobserved) trajectory function; $\sigma^{2}$ is the error variance; ${\bm \beta}$ is a $q_{1}$-dimensional fixed effects vector and ${\bm b}_{i}$ is an $r$-dimensional random effects vector independently distributed according to a normal distribution with a mean at zero and variance-covariance matrix ${\bm \Sigma}_{r \times r}$. Denote ${\bm\theta}^{^L}=({\bm \beta},\sigma^{2},{\bm\Sigma}_{r \times r})$. It is common to assume that $\mu_{i}(t \mid {\bm b}_{i}, {\bm \beta})$ takes a linear form:
\begin{equation*}
    \mu_{i}(t \mid {\bm b}_{i}, {\bm \beta}) = X^{T}_{i}(t){\bm \beta}+Z^{T}_{i}(t){\bm b}_{i},
\end{equation*}

\noindent where $X_{i}(t)$ and $Z_{i}(t)$ are the possibly time-dependent design vectors for the fixed and random effects, respectively. In this case, \eqref{eq:REM} specifies a standard linear mixed model (LMM).
More general types of markers (e.g., binary count variables) can be coped with by considering a (multivariate) generalized linear mixed model.

\subsection{Multistate submodel}

Let $\{E_{i}(t),t\geq 0\}$ be a multistate process for subject $i$ with a common state space $S=\{0,1,\ldots,N\}$, where $E_{i}(t)$ denotes the state that subject $i$ occupies at time $t$. 
The law of the process $E_{i}(t)$ can be fully characterized by the transition intensities between the two states
\begin{equation}\label{eq:def_MSM}
h^{(i)}_{jk}(t;\mathcal{H}_{t^{-}})=\lim_{\Delta t\to 0}\frac{P(E_{i}(t+\Delta t)=k \mid E_{i}(t)=j; \mathcal{H}_{t^{-}})}{\Delta t}, \quad j\neq k\in S, 
\end{equation}

\noindent which represents the instantaneous risk of transiting from state $j$ to state $k$ at time $t$ given the history up to time $t$, $\mathcal{H}_{t^{-}}$.
Here we make the simplifying assumption that $h_{jk}^{(i)}(t;\mathcal{H}_{t^{-}})=h_{jk}^{(i)}(B(t))$, where $B(t)$ is the time since entry into the current state $j$. This results in what is known as a clock-reset semi-Markov MSM as the time scale of the model is reset to zero after entering a new state. 
The regression models for the transition intensities are specified as
\begin{equation}\label{eq:intensity_MSM}
    h_{jk}^{(i)}(B(t) \mid {\bm\theta}^{^E}_{jk})=h_{0,jk}(B(t) \mid \phi_{jk})\exp(w^{T}_{i}\gamma_{jk}+g({\bm \beta},{\bm b}_{i},t,\alpha_{jk})),
\end{equation}

\noindent with $h_{0,jk}(\cdot)$ representing a baseline hazard function parameterized by $\phi_{jk}$, $w_{i}$ be the $q_{2}$-dimensional vector of exogenous risk factors associated with the coefficient vector $\gamma_{jk}$ and ${\bm\theta}^{^E}_{jk}=(\phi_{jk},\gamma_{jk},\alpha_{jk})$.
The function $g$, parametrized by $\alpha_{jk}$, describes the association between the longitudinal marker's dynamics and the multistate process.
Some common choices of $g$ are $g=\alpha_{jk}\mu_{i}(t)$ (current value association), $g=\alpha_{jk}\mu_{i}^{'}(t)$ (current slope association) or $g=\alpha_{jk}^{T}{\bm b}_{i}$ (shared random effects). In particular, $\alpha_{jk}$ quantifies the strength of the association between the two processes. Of course, depending on the context of the application, other reasonable functional forms can be used here.
Note that not every transition intensity in \eqref{eq:def_MSM} needs to be modelled. According to the multistate transition diagram, which is prefixed by design or ground knowledge, some of the intensities will be identical to zero. For instance, in the multistate diagram shown in Figure \ref{fig:example}, there are only six permitted direct transitions and thus only six transition models need to be considered.

To formally define the model likelihood, we assume that each subject is followed up continuously for some period of time, subject to a right censoring time $C_{i}$, and let $E_{i}=\{E_{i}(t), 0\leq t\leq C_{i}\}$ be the observed process.
Let $(T^{(i)}_{1},\ldots,T^{(i)}_{N_{i}})$ be the sequence of observed transition times such that $T^{(i)}_{l}<T^{(i)}_{l+1}$; $N_{i}$ is the total number of observed transitions ($N_{i}=0$ if the subject is censored without having a transition), and let $D^{(i)}_{l}=T^{(i)}_{l}-T^{(i)}_{l-1}$, $l=1,\ldots,N_{i}+1$, where $T^{(i)}_{0}=0$ and $T^{(i)}_{N_{i}+1}=C_{i}$, be the sequence of sojourn times.
We additionally define the transition indicator variables $\delta^{(i)}_{l,jk}$ such that $\delta^{(i)}_{l,jk}=1$ if the transition $j\to k$ occurs at event time $T^{(i)}_{l}$ and $\delta^{(i)}_{l,jk}=0$ otherwise. With the notation and assumptions above, the likelihood contribution of the multistate process for subject $i$ can be expressed as
\begin{equation}\label{eq:likMSM}
f(E_{i} \mid \cdot)=\prod_{1\leq l \leq N_{i}+1}\biggl(\prod_{j\neq k\in S}h_{jk}^{(i)}(D^{(i)}_{l})^{\delta^{(i)}_{l,jk}}\biggl)\exp\Bigl(-\int_{0}^{D^{(i)}_{l}}\sum_{j\in S}h_{E_{i}(T^{(i)}_{l-1}),j}^{(i)}(u)du\Bigl),
\end{equation}

\noindent where the intensity functions $h_{jk}^{(i)}(\cdot)$ are given by \eqref{eq:intensity_MSM} (for brevity, here and in what follows we omit the conditioned parameter set) and we adopt the convention that $0^{0}$ is taken to be one. A rigorous derivation of the likelihood can be made based on the counting process theory, see \citet{cook2018multistate} for details. Note that in general integrals in \eqref{eq:likMSM} cannot be computed analytically and numerical approximation is required. Gauss quadrature methods are the standard options for their evaluation and in our implementation, we used the Gaussian-Legendre quadrature method (with 15 quadrature points).

\subsection{Bayesian inference}
In Bayesian inference, our interest lies in the posterior distribution of model parameters and random effects, which is assumed to be factorised as
\begin{equation}\label{eq:Post_JMSM}
\begin{gathered} 
    f(\Theta, {\bm b} \mid \text{data})\propto \mathcal{L}(\Theta,{\bm b}\mid \text{data})f({\bm b} \mid \Theta)f(\Theta),\\
    = \Bigl(\prod_{i=1}^{n}\prod_{j=1}^{n_{i}}f(y_{ij} \mid b_{i},\Theta)f(E_{i} \mid b_{i},\Theta)\Bigl)\Bigl(\prod_{i=1}^{n}f({\bm b_{i}} \mid \Theta)\Bigl)f(\Theta),
\end{gathered}
\end{equation}

\noindent where $\Theta=({\bm\theta}^{^L},{\bm\theta}^{^E})$ is the full model parameter vector with ${\bm\theta}^{^E}=\{\theta^{E}_{jk}\}$ and ${\bm b}=({\bm b}_{1},\ldots,{\bm b}_{n})$. The conditional densities in \eqref{eq:Post_JMSM} are given in \eqref{eq:REM} and \eqref{eq:likMSM}. As commonly adopted in the joint model literature, here we make the simplifying assumption that the longitudinal and multistate processes are conditionally independent given the random effects \citep{rizopoulos2012joint,ferrer2016joint}. 
To complete the Bayesian formulation, prior distributions need to be chosen for the parameters in the model. In practice, independent and weakly-informative prior distributions are commonly used for joint models \citep{papageorgiou2019overview, alvares2021tractable}. More specifically, independent diffuse Normal priors centered at zero are typically assumed for each of the fixed effects $\beta$, $\gamma_{jk}$ and the association parameters $\alpha_{jk}$; in scenarios where multiple covariates and/or association structures are being included shrinkage priors may be adopted, see e.g. \citet{andrinopoulou2016bayesian}.
 For other parameters, we could use an inverse Gamma prior for the error variance and an inverse-Wishart prior for the random effects covariance matrix. Once the parametric form of the baseline transition intensity functions $h_{0,jk}(u \mid \phi_{jk})$ are defined, weakly-informative priors can also be specified for their parameters.

The posterior defined in \eqref{eq:Post_JMSM} is not analytically tractable. For this paper, we use the No-U-Turn Sampler (NUTS) \citep{hoffman2014no}, an adaptively tuned auxiliary variable MCMC algorithm that exploits the Hamiltonian dynamics \citep{neal2011mcmc}, to sample from the targeted posterior distribution. The NUTS often offers much better sampling efficiency than other standard MCMC algorithms such as the random walk Metropolis-Hastings and Gibbs samplers, especially when the parameters are of high dimension and complex correlation structure \citep{betancourt2017conceptual}.
Note that we make no claim that NUTS is the best sampler for the model of study, but we find its performance to be satisfactory for our purpose.

\section{Blockwise parallel inference} \label{sec:block}
In this section, we describe the proposed intuitive blockwise inference approaches to alleviate the computational challenges of fitting the joint longitudinal and multistate model described in Section~\ref{sec:jm-msm}. Further technical details on the asymptotic and finite sample comparisons of the inference approaches are provided in the appendix. To help illustrate the idea, we employ a toy example throughout as shown in Figure \ref{fig:example}, where the event process is depicted by a 6-state progressive multistate process and we have a single continuous longitudinal outcome.

\subsection{The competing risks approach}
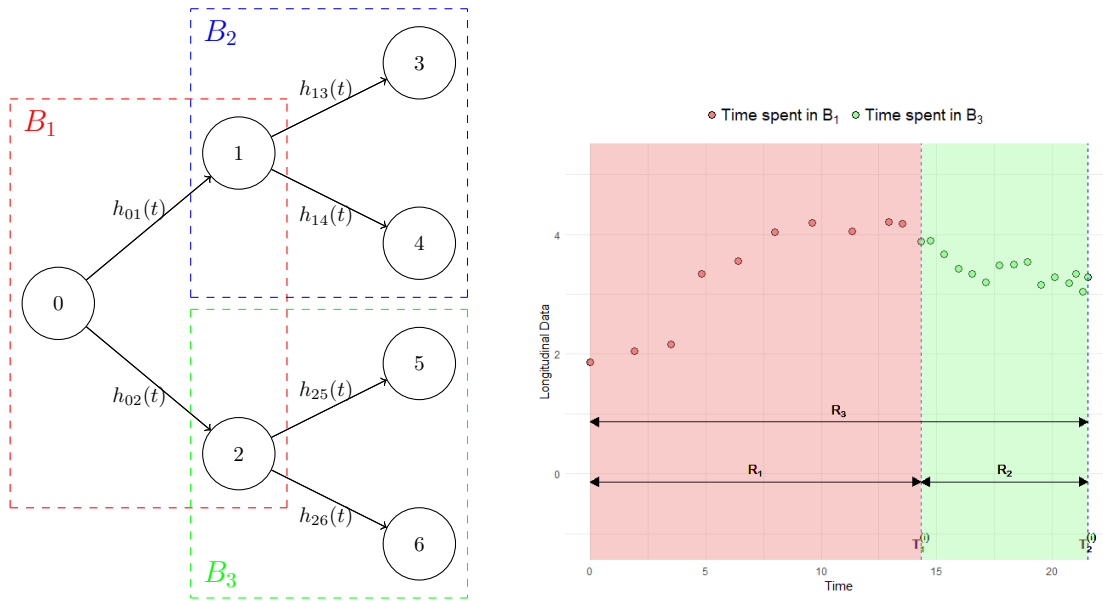
\begin{figure}[htbp]
    \centering
    \begin{minipage}[b]{0.5\textwidth}
        \centering
        \begin{tikzpicture}[scale=0.8, transform shape]
            \tikzstyle{sty-01} = [circle, draw]
        \node[sty-01] (n0) at (-4.0,+0.0) 
        {$\quad 0 \quad$};
        \node[sty-01] (n1) at (-1.0,+2.5) 
        {$\quad 1 \quad$};
        \node[sty-01] (n2) at (-1.0,-2.5) 
        {$\quad 2 \quad$};
        \node[sty-01] (n3) at (+2.0,+4.0) 
        {$\quad 3 \quad$};
        \node[sty-01] (n4) at (+2.0,+1.0) 
        {$\quad 4 \quad$};
        \node[sty-01] (n5) at (+2.0,-1.0) 
        {$\quad 5 \quad$};
        \node[sty-01] (n6) at (+2.0,-4.0) 
        {$\quad 6 \quad$};
        \draw[->] (n0) -> (n1);
        \draw[->] (n0) -> (n2);
        \draw[->] (n1) -> (n3);
        \draw[->] (n1) -> (n4);
        \draw[->] (n2) -> (n5);
        \draw[->] (n2) -> (n6);
        \draw[->] (n0) -- node[anchor=south] {$h_{01}(t) \;\;\;$} (n1);
        \draw[->] (n0) -- node[anchor=north] {$h_{02}(t) \;\;\;$} (n2);
        \draw[->] (n1) -- node[anchor=south] {$h_{13}(t) \;$} (n3);
        \draw[->] (n1) -- node[anchor=north] {$h_{14}(t) \;$} (n4);
        \draw[->] (n2) -- node[anchor=south] {$h_{25}(t) \;$} (n5);
        \draw[->] (n2) -- node[anchor=north] {$h_{26}(t) \;$} (n6);
        \draw (-4.3,3.0) node {\Large\textcolor{red}{$B_{1}$}};
        \draw[red, dashed] (-4.8,-3.4)--(-4.8,3.4);
        \draw[red, dashed] (-4.8,-3.4)--(-0.2,-3.4);
        \draw[red, dashed] (-4.8,3.4)--(-0.2,3.4);
        \draw[red, dashed] (-0.2,-3.4)--(-0.2,3.4);
        \draw (-1.3,4.5) node {\Large\textcolor{blue}{$B_{2}$}};
        \draw[blue, dashed] (-1.8,0.1)--(-1.8,4.9);
        \draw[blue, dashed] (-1.8,0.1)--(2.8,0.1);
        \draw[blue, dashed] (-1.8,4.9)--(2.8,4.9);
        \draw[blue, dashed] (2.8,0.1)--(2.8,4.9);
        \draw (-1.3,-4.5) node {\Large\textcolor{green}{$B_{3}$}};
        \draw[green, dashed] (-1.8,-0.1)--(-1.8,-4.9);
        \draw[green, dashed] (-1.8,-0.1)--(2.8,-0.1);
        \draw[green, dashed] (-1.8,-4.9)--(2.8,-4.9);
        \draw[green, dashed] (2.8,-0.1)--(2.8,-4.9);
        \end{tikzpicture}
    \end{minipage}%
    \hfill
    \begin{minipage}[b]{0.5\textwidth}
        \centering
       \includegraphics[width=\textwidth]{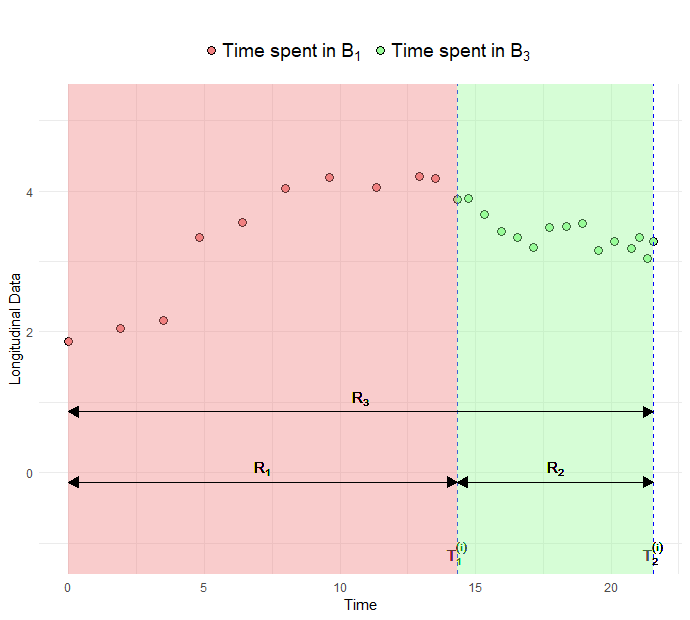}
    \end{minipage}
    \caption{Illustration of the blockwise inference approach using a toy example. Left panel: multistate progressive scheme with its transition blocks. For the JM-CR approach there are 3 blocks: $B_{1}$, $B_{2}$ and $B_{3}$, indicated by colored dashed boxes. The JM-ST approach has 6 blocks, corresponding to each permitted transition: $0\to 1$, $0\to 2$, $1\to 3$, $1\to 4$, $2\to 5$ and $2\to 6$, with respective transition intensities $h_{01}(t)$, $h_{02}(t)$, $h_{13}(t)$, $h_{14}(t)$, $h_{25}(t)$ and $h_{26}(t)$.
    Right panel: hypothetical longitudinal profile of a marker over time for an imaginary subject experiencing transition $0\to 2$ at $T_{1}^{(i)}$ and transition $2\to 5$ at $T_{2}^{(i)}$. The data collected during the time spent in $B_{1}$ (indicated by $R_{1}$) and $B_{2}$ (indicated by $R_{2}$) are indicated with red and green shades, respectively. The entire follow-up period is indicated by $R_{3}$. }
    \label{fig:example}
\end{figure}

Our initial proposal involves breaking down the (clock-reset) Markov multistate process into separate connected blocks of competing risk processes. Based on this decomposition, we can simultaneously and independently fit a joint longitudinal and competing risk (JM-CR) model to each competing risk block, using only the longitudinal and time-to-event data that are associated with the block. In our toy example (see Figure \ref{fig:example}), the multistate process can be decomposed into three competing risk blocks, $B_{1}$, $B_{2}$ and $B_{3}$, each consisting of an initial state and two absorbing states (the definitions of initial and absorbing states are relative to their respective blocks; e.g. states 1 and 2 are absorbing states of $B_{1}$ but are treated as initial states in subsequent blocks). Consider a generic competing risk block $B_{v}$. Let $\mathcal{I}_{B_{v}}$ denote the index set of subjects who entered the initial state in $B_{v}$ and were at risk for transitions in $B_{v}$. Let $\mathcal{T}^{(i)}_{B_{v}}$ denote the index set for time point $j$ such that $\{y_{ij}\}_{j\in \mathcal{T}^{(i)}_{B_{v}}}$ represents the longitudinal data linked to block $B_{v}$ for subject $i$ (see Section \ref{subsec:linklongitudinal} for more details on the construction of this index set), $E^{B_{v}}_{i}$ be the observed competing risk process associated with $B_{v}$, and $S_{B_{v}}$ be the set of the indices of the absorbing state in $B_{v}$. Then the likelihood function for the JM-CR approach for $B_{v}$ can be expressed as 
\begin{equation}\label{eq:likJCR}
\mathcal{L}(\Theta_{B_{v}},{\bm b_{B_{v}}}\mid \text{data})=\prod_{i\in \mathcal{I}_{B_{v}}}\prod_{j\in \mathcal{T}^{(i)}_{B_{v}}}f(y_{ij} \mid b_{i},\Theta_{B_{v}})f(E^{B_{v}}_{i} \mid b_{i},\Theta_{B_{v}}),
\end{equation}
where $\Theta_{B_{v}}$ and $\bm b_{B_{v}}$ are vectors of the associated model parameters and random effects, and
\begin{equation}\label{eq:likCR}
    f(E^{B_{v}}_{i} \mid \cdot) = \prod_{k\in S_{B_{v}}}h_{j_{v}k}^{(i)}(D^{(i)})^{\delta^{(i)}_{j_{v}k}}\exp\Bigl(-\int_{0}^{D^{(i)}}\sum_{l\in S_{B_{v}}}h_{j_{v},l}^{(i)}(u)du\Bigl),
\end{equation}
where $j_{v}$ denotes the initial state index in $B_{v}$. $\delta^{(i)}_{jk}$ is the transition indicator variable defined analogously as before, indicating if the individual makes a transition from $j\to k$: $D^{(i)}=T^{(i)}$ if a transition is observed at $T^{(i)}$ and else $D^{(i)}=C_{i}$, the right censoring time. $\delta^{(i)}_{jk}$, $T^{(i)}$ and $C_{i}$ are all defined with respect to the block $B_{v}$, with the time origin being reset after individual $i$ entered $B_{v}$.

The gain in computational efficiency of working with the JM-CR approach is clear when comparing the likelihood in \eqref{eq:likJCR} to that in \eqref{eq:Post_JMSM}: the computation of the block-based joint likelihood involves only a subset of subjects' data that is relevant to the block, and the likelihood for a competing risk process (given in \eqref{eq:likCR}) is less costly to evaluate than the multistate counterpart (given in \eqref{eq:likMSM}) - the latter, in fact, takes the form of a product of likelihood for the competing risk process in each block. Furthermore, sampling from $(\Theta_{B_{v}},{\bm b_{B_{v}}})$ independently for each block, instead of sampling from the much higher dimensional space of $(\Theta, {\bm b})$ all at once, could lead to an additional gain in sampling efficiency (e.g. faster convergence of the Markov chain).

\subsection{The single transition approach}
When the focus is on estimating the association and other MSM parameters, we can further decompose the task of estimating a JM-CR model into estimating a joint longitudinal and survival model independently for each allowed transition $i\to j$ within the block, which we shall refer to as the JM-ST approach.
This approach is motivated by the factorization property of the CR likelihood
\begin{equation}\label{eq:likUS}
    f(E^{B_{v}}_{i} \mid \cdot) = \prod_{k\in S_{B_{v}}}h_{j_{v}k}^{(i)}(D^{(i)})^{\delta^{(i)}_{j_{v}k}}\exp\Bigl(-\int_{0}^{D^{(i)}}h_{j_{v},k}^{(i)}(u)du\Bigl) = \prod_{k\in S_{B_{v}}}f(E^{B_{j_{v},k}}_{i} \mid \cdot),
\end{equation}
where $E^{B_{j_{v},k}}_{i}$ denotes the observed survival process for subject $i$ considering the transition $j_{v}\to k$, $k\in S_{B_{v}}$ (i.e. subjects who transition to a competing absorbing state or are censored are both treated as censored), and $f(E^{B_{j_{v},k}}_{i} \mid \cdot)$ takes the form of a standard survival likelihood.
The working likelihood function of the JM-ST approach for the single-transition block $B_{j_{v},k}$ is given by
\begin{equation}\label{eq:likJUS}
\mathcal{L}(\Theta_{B_{j_{v},k}},{\bm b_{B_{j_{v},k}}}\mid \text{data})=\prod_{i\in \mathcal{I}_{B_{v}}}\prod_{j\in \mathcal{T}^{(i)}_{B_{v}}}f(y_{ij} \mid b_{i},\Theta_{B_{j_{v},k}})f(E^{B_{j_{v},k}}_{i} \mid b_{i},\Theta_{B_{j_{v},k}}),
\end{equation}
where $\Theta_{B_{j_{v},k}}$ and ${\bm b_{B_{j_{v},k}}}$ are vectors of the associated model parameters and random effects, the index sets $\mathcal{I}_{B_{v}}$ and $\mathcal{T}^{(i)}_{B_{v}}$ are defined as in the JM-CR approach, and $f(E^{B_{j_{v},k}}_{i} \mid \cdot)$ is given in \eqref{eq:likUS}. 
This implies that when using the JM-ST approach to estimate each transition, $B_{j_{v},k}$, the same subset of subjects ($\mathcal{I}_{B_{v}}$) is used as in the JM-CR approach for jointly estimating block $B_{v}$, and the same set of longitudinal data is used if the same strategy for linking the longitudinal data to the block ($\mathcal{T}^{(i)}_{B_{v}}$) is adopted.
The computational advantage of working with JM-ST over JM-CR stems from the fact that the likelihood of the associated event process is simpler to compute for the former than for the latter, making it overall less computationally costly. Sampling with JM-ST could also be more efficient due to a further reduction in the number of parameters to be jointly sampled. However, note that, unlike the JM-MSM or JM-CR approaches, individual-specific longitudinal trajectories cannot be estimated using this single transition approach, since the subjects are reused for estimating each transition within the block. This results in more than one set of estimates for the longitudinal model parameters and random effects for each subject involved in the CR block.

\subsection{Incorporating longitudinal data in model blocks}
\label{subsec:linklongitudinal}
When applying the JM-CR or JM-ST approach to perform inference in a specific block, it is necessary to specify the index set for the longitudinal data points, $\mathcal{T}^{(i)}_{B_{v}}$. This index set determines which longitudinal data will be included in the likelihood function. Depending on the block and model structure, there could be various ways to define this index set. In this paper, we consider two natural strategies. 

The first strategy involves using only concurrent longitudinal measurements collected during the time subjects spent in the given block. In the example shown in Figure \ref{fig:example}, this would correspond to using only the red (collected during $R_{1}$) or green (collected during $R_{2}$) shaded data points for estimating blocks 1 or 3. In cases where no longitudinal data exists within the block, strategies such as the last observation carry forward can be employed to impute a value at the time entering the current state.
This approach inherently allows for the handling of potential changes in longitudinal dynamics due to the experience of an event by using only the ``local" information, thus enabling more adaptive modelling of the marker's dynamics. Another strategy is to incorporate all historical data collected since the start of the process to improve estimation, which would be particularly relevant when most subjects have very limited longitudinal data points within the block. Referring again to the example shown in Figure \ref{fig:example}, this means that when estimating $B_{3}$, both red and green shaded longitudinal data points (collected during $R_{3}$) will be used in constructing the longitudinal submodel. However, in this hypothetical scenario, a standard linear mixed model may not be able to adequately capture the structural change in the longitudinal dynamics. A more formal and systematic comparison between these two options can be conducted within the Bayesian inferential framework in a blockwise manner, as will be presented in Section \ref{subsec:modelcompare}.

\subsection{Model comparison}
\label{subsec:modelcompare}
To evaluate or compare candidate models for the same data, it is common to rely on the measure of expected out-of-sample predictive ability. However, this measure cannot be directly computed as the true data-generating process is unknown. Here, we consider a Bayesian leave-one-out cross-validation (LOO-CV) score as suggested in \citet{vehtari2017practical} for approximating the predictive accuracy of a fitted Bayesian model. The LOO-CV score is defined as
\begin{equation}\label{eq:loocv}
   \text{LOO-CV}=\sum_{i=1}^{n}\log f(D_{i} \mid D^{(n)}_{-i}),
\end{equation}
where $D_{i}$ denotes the $i$th observation, $D^{(n)}=(D_{1},\ldots,D_{n})$ and $D^{(n)}_{-i}$ represents the data set excluding the $i$th observation. Note that $f(D_{i} \mid D^{(n)}_{-i})=\int f(D_{i} \mid \theta,D^{(n)}_{-i})f(\theta \mid D^{(n)}_{-i})d\theta$, where $\theta$ is a vector of model parameters, measures how well the model would predict the $i$th observation based on the model estimated using the data without including that observation. The LOO-CV can be effectively estimated based on existing simulated MCMC samples using a Pareto smoothed importance sampling approach proposed in \citet{vehtari2017practical}, which is implemented in the R package loo \citep{loo}, avoiding the need to refit the model $n$ times.
As an alternative to cross-validation, the widely applicable information criterion (WAIC) can also be used for estimating the out-of-sample prediction accuracy. WAIC is asymptotically equivalent to the LOO-CV and has advantages over other predictive criteria, such as the deviance information criterion (DIC). However, the authors in \citet{vehtari2017practical} noted that WAIC can be less robust than LOO-CV in finite cases with weak priors. In our applications, we utilized LOO-CV.

As previously discussed, a practical issue when implementing the JM-CR or JM-ST approach is determining the extent of longitudinal information to use for a given block. This choice involves e.g. either utilizing all historical longitudinal data or using only concurrent data collected within the block. As we will demonstrate later, this decision can impact estimation accuracy in some scenarios. To address this issue, we suggest fitting the model with both versions and selecting the more ``appropriate" one based on the LOO-CV computed for the data within the block. Note that in this joint modelling context, different definitions of ``a single data point" exist. For instance, the $i$th observation $D_{i}$ could be defined as i) a collection of a subject's longitudinal data, $y_{i}$, or a single measurement $y_{ij}$; ii) the observed time-to-event data $E_{i}^{B_{v}}$ (for JM-CR) or $E^{B_{j_{v},k}}_{i}$ (for JM-ST); or iii) both longitudinal and time-to-event data, i.e. $D_{i}=(y_{i},E_{i}^{B_{v}})$ (for JM-CR) or $D_{i}=(y_{i},E^{B_{j_{v},k}}_{i})$ (for JM-ST). Here as our focus lies on the longitudinal component, we suggest using either version i) or iii). In our experience, both versions yield consistent results. Depending on the model comparison context, different versions may be used.

\section{Simulation study} \label{sec:sim}
We conducted a simulation study to evaluate and compare the accuracy and computational efficiency of the candidate approaches for inference in Bayesian joint multistate models under both hypothetical and realistic scenarios. 
For our proposed JM-CR and JM-ST approaches, we considered two different configurations: one using only concurrent longitudinal data (JM-CR-C and JM-ST-C), and the other using all historical longitudinal data (JM-CR-H and JM-ST-H), both in relation to the block under consideration. The benchmark comparator is the standard joint multistate approach as described in Section \ref{sec:jm-msm} (JM-MSM). For all approaches, posterior simulation is performed using RStan version 2.21.7 \citep{Rstan}, a probabilistic programming environment that provides an implementation of NUTS. The R and Stan codes will be made available at \url{http://www.github.com/sidachen55/Blockwise_JM_MSM}.

\subsection{Simulation models} \label{subsec:sim_models}
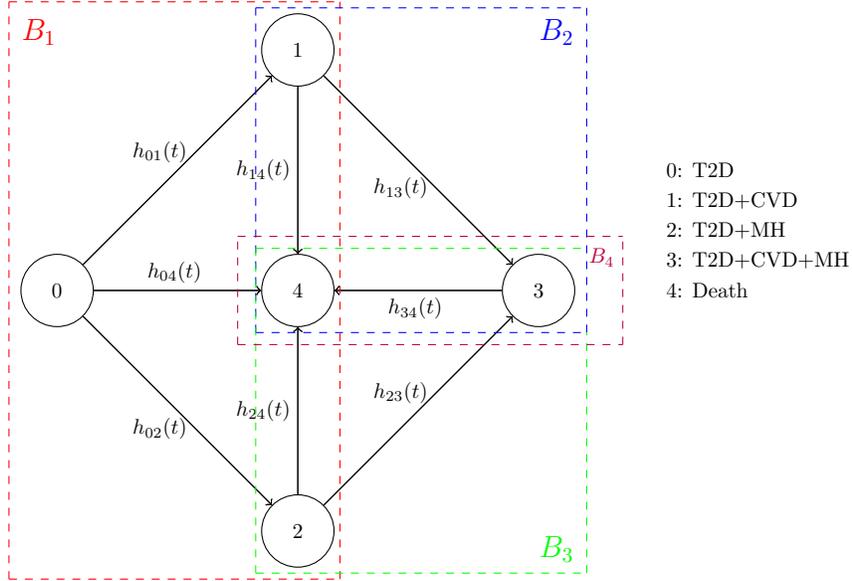
\begin{figure}[h!]
    \centering
    \begin{tikzpicture}[scale=0.8, transform shape]
        \tikzstyle{sty-01} = [circle, draw]
        \node[sty-01] (n0) at (-4.0,+0.0) 
        {$\quad 0 \quad$};
        \node[sty-01] (n1) at (-0.0,+4.0) 
        {$\quad 1 \quad$};
        \node[sty-01] (n2) at (-0.0,-4.0) 
        {$\quad 2 \quad$};
        \node[sty-01] (n3) at (+4.0,+0.0) 
        {$\quad 3 \quad$};
        \node[sty-01] (n4) at (-0.0,+0.0) 
        {$\quad 4 \quad$};
        \draw[->] (n0) -> (n1);
        \draw[->] (n0) -> (n2);
        \draw[->] (n0) -> (n4);
        \draw[->] (n1) -> (n3);
        \draw[->] (n1) -> (n4);
        \draw[->] (n2) -> (n3);
        \draw[->] (n2) -> (n4);
        \draw[->] (n3) -> (n4);
        \draw[->] (n0) -- node[anchor=south] {$h_{01}(t) \;\;\;\;\;\;$} (n1);
        \draw[->] (n0) -- node[anchor=north] {$h_{02}(t) \;\;\;\;\;\;$} (n2);
        \draw[->] (n0) -- node[anchor=south] {$h_{04}(t) \;$} (n4);
        \draw[->] (n1) -- node[anchor=north] {$h_{13}(t) \;\;\;\;\;\;$} (n3);
        \draw[->] (n1) -- node[anchor=east] {$h_{14}(t) $} (n4);
        \draw[->] (n2) -- node[anchor=south] {$h_{23}(t) \;\;\;\;\;\;$} (n3);
        \draw[->] (n2) -- node[anchor=east] {$h_{24}(t) $} (n4);
        \draw[->] (n3) -- node[anchor=north] {$h_{34}(t) \;$} (n4);
        \draw (-4.3,4.3) node {\Large\textcolor{red}{$B_{1}$}};
        \draw[red, dashed] (-4.8,-4.8)--(-4.8,4.8);
        \draw[red, dashed] (-4.8,-4.8)--(0.7,-4.8);
        \draw[red, dashed] (-4.8,4.8)--(0.7,4.8);
        \draw[red, dashed] (0.7,-4.8)--(0.7,4.8);
        \draw (4.3,4.3) node {\Large\textcolor{blue}{$B_{2}$}};
        \draw[blue, dashed] (-0.7,-0.7)--(-0.7,4.7);
        \draw[blue, dashed] (-0.7,-0.7)--(4.8,-0.7);
        \draw[blue, dashed] (-0.7,4.7)--(4.8,4.7);
        \draw[blue, dashed] (4.8,-0.7)--(4.8,4.7);
        \draw (4.3,-4.3) node {\Large\textcolor{green}{$B_{3}$}};
        \draw[green, dashed] (-0.7,-4.7)--(-0.7,0.7);
        \draw[green, dashed] (-0.7,0.7)--(4.8,0.7);
        \draw[green, dashed] (-0.7,-4.7)--(4.8,-4.7);
        \draw[green, dashed] (4.8,-4.7)--(4.8,-0.7);
        \draw (5.05,0.55) node {\textcolor{purple}{$B_{4}$}};
        \draw[purple, dashed] (-1.0,-0.9)--(-1.0,0.9);
        \draw[purple, dashed] (-1.0,0.9)--(5.4,0.9);
        \draw[purple, dashed] (-1.0,-0.9)--(5.4,-0.9);
        \draw[purple, dashed] (5.4,-0.9)--(5.4,0.9);

    \begin{scope}[shift={(6,2)}]
    \node[right] at (0,0) {0: T2D};
    \node[right] at (0,-0.5) {1: T2D+CVD};
    \node[right] at (0,-1) {2: T2D+MH};
    \node[right] at (0,-1.5) {3: T2D+CVD+MH};
    \node[right] at (0,-2) {4: Death};
    
    \end{scope}
    \end{tikzpicture}
    \caption{Multistate progressive scheme and its transition blocks. A total of eight different transitions are allowed: $0\to 1$, $0\to 2$, $0\to 4$, $1\to 3$, $1\to 4$,  $2\to 3$, $2\to 4$ and $3\to 4$ with corresponding transition intensities $h_{01}(t)$, $h_{02}(t)$, $h_{04}(t)$, $h_{13}(t)$, $h_{14}(t)$, $h_{23}(t)$, $h_{24}(t)$ and $h_{34}(t)$. $B_{1}$ to $B_{4}$ are the 4 transition blocks for the JM-CR approach. The JM-ST approach has 8 blocks, corresponding to each permitted transition. In our application context, states symbolize different disease states as indicated in the legend, where T2D denotes type-2 diabetes, CVD denotes cardiovascular diseases and MH denotes mental health conditions. \label{fig:mstate2}}
\end{figure}

We considered two different simulation models, which are briefly outlined below. 
Model 1 is based on our toy example shown in Figure \ref{fig:example}. For the multistate process, the transition intensities are specified as  
\begin{equation}\label{eq:sim_intensity_MSM}
    h_{jk}^{(i)}(B(t))=h_{0,jk}(B(t))\exp(w_{i}\gamma_{jk}+\alpha_{jk}\mu_{i}(t)),
\end{equation}
where $(j,k)\in \{(0,1),(0,2),(1,3),(1,4),(2,5),(2,6)\}$, $h_{0,jk}(u)=\delta_{jk}u^{\delta_{jk}-1}\lambda_{jk}$ is the Weibull hazard function specified by the shape parameter $\delta_{jk}$ and scale parameter $\lambda_{jk}$, and $w_{i}$ is a one dimensional hypothetical baseline covariate for subject $i$ and $\mu_{i}(t)$ is the unobserved longitudinal trajectory for subject $i$ at time $t$.
For the longitudinal process, we consider three different data-generating scenarios. In both scenarios 1 and 2, the longitudinal data is generated from a basic linear mixed model with a random intercept and slope:
\begin{equation}\label{eq:sim_LMM}
y_{i}(t)=\mu_{i}(t)+\epsilon_{i}(t), \quad \epsilon_{i}(t)\sim N(0,\sigma_{e}^{2}),
\end{equation}
with $\mu_{i}(t)=\beta_{1}+b_{i1}+(\beta_{2}+b_{i2})t$ and $(b_{i1},b_{i2})\sim N(0,\Sigma)$.
For the visiting process, i.e. the mechanism that generates the time points $t_{ij}$ at which longitudinal data is collected, we set the distance between the $t_{ij}$ to be equidistant with an interval of $\Delta_{1}$ when the subject stays in $B_{1}$ and is switched to $\Delta_{2}$ after the subject experiences the first transition (i.e. enters later blocks).
The difference between the 2 scenarios lies in the sampling frequency. For scenario 1, we set $\Delta_{1} > \Delta_{2}$ so that the data is more densely collected after the first transition, whereas in scenario 2, $\Delta_{1} < \Delta_{2}$. This allows us to examine the effect of the number of longitudinal measurements per subject on the inference.
In scenario 3, we create a structural change in the underlying longitudinal trajectory, which occurs immediately after the first transition is experienced, i.e. 
\begin{equation}\label{eq:sim_LMM2}
    \mu_{i}(t)=\beta_{1}^{'}+b_{i1}+(\beta_{2}^{'}+b_{i2})t, \quad t\geq T^{(i)}_{1}, 
\end{equation}
where $\beta_{1}^{'}\neq \beta_{1}$ and $\beta_{2}^{'}\neq \beta_{2}$. We believe that this situation is likely to happen in certain cases where changes may be attributed to changes in the underlying status of the subject or external intervention. The random intercept and slope could change as well but here we keep them fixed. For this scenario, the same visiting process as in scenario 1 is used. Our second simulation model, Model 2, is motivated by a slightly more complex and realistic setting encountered in the analysis of longitudinal multimorbidity progression (see our case study in Section \ref{sec:appl}). The associated multistate transition diagram is shown in Figure \ref{fig:mstate2}.
We can identify 4 blocks for the JM-CR approach and 8 blocks for the JM-ST approach, noting that block $B_{4}$ is shared for both approaches.
The regression models for the transition intensities and the longitudinal data are specified as in Model 1 (scenario 1). For the visiting process, let the sampling interval after the second transition (i.e. entering block $B_{4}$) be denoted by $\Delta_3$, we set $\Delta_{1} > \Delta_{2} > \Delta_{3}$. A comparison of the simulation models considered is summarized in Table \ref{tab:compare-sim-models}.

\begin{table}[h]
\centering
\caption{Comparison of simulation models.}
\begin{tabular}{l c c c}
\toprule
\textbf{Simulation model} & \textbf{MSM structure} & \textbf{LMM setting} & \textbf{Visiting process} \\
\midrule
Model 1 (scenario 1) & Figure \ref{fig:example} & Equation \eqref{eq:sim_LMM} & $\Delta_{1}>\Delta_{2}$ \\
Model 1 (scenario 2) & Figure \ref{fig:example} & Equation \eqref{eq:sim_LMM} & $\Delta_{1}<\Delta_{2}$ \\
Model 1 (scenario 3) & Figure \ref{fig:example} & Equations \eqref{eq:sim_LMM} and \eqref{eq:sim_LMM2} & $\Delta_{1}>\Delta_{2}$ \\
Model 2  & Figure \ref{fig:mstate2} & Equation \eqref{eq:sim_LMM} & $\Delta_{1}>\Delta_{2}>\Delta_{3}$ \\
\bottomrule
\end{tabular}
\label{tab:compare-sim-models}
\end{table}

Given the parameter values, a joint longitudinal and multistate process can be simulated as follows. Firstly, subject-specific random effects $b_{i}$ are simulated from its prior distribution $N(0, \Sigma)$. Conditional on the $b_{i}$, for each subject $i=1,\ldots,n$, we generate a right censoring time point $C_{i}$ (either randomly or deterministically), and then simulate the multistate process $E_{i}(t)$ for $0\leq t\leq C_{i}$ using the scheme as described in e.g. \citet{crowther2017parametric}. Finally, the longitudinal trajectory is simulated by first generating the measurement time points $t_{ij}$ (either randomly or deterministically) within the follow-up period $0\leq t\leq C_{i}$, and then generating the data $y_{i}(t_{ij})$ according to the specified longitudinal submodel.

\subsection{Parameter settings}
We use the following parameter settings to simulate data from the simulation models.
For Model 1, random effects $b_{1},\ldots,b_{n}$ are generated according to a bivariate normal distribution $N(\big(\begin{smallmatrix}
  0 \\
  0 
\end{smallmatrix}\big),\big(\begin{smallmatrix}
  \sigma_{1}^{2} & \rho\sigma_{1}\sigma_{2}\\
  \rho\sigma_{1}\sigma_{2} & \sigma_{2}^{2}
\end{smallmatrix}\big))$ with $\sigma_{1}=0.4$, $\sigma_{2}=0.3$ and $\rho=0.4$. For the multistate process, we set $\gamma_{jk}=1$, $\alpha_{jk}=0.9$, $\delta_{jk}=1.3$ and $\lambda_{jk}=\exp(-6)$ for all transitions. $w_{i}$ is considered as an artificial ``age" covariate, generated from a mixture of uniform distributions defined over the intervals $(18,65)$, $(65,80)$ and $(80,90)$ with weights $0.45$, $0.3$ and $0.25$, respectively. The variable is then normalized to have a mean of 0 and a standard deviation of 1.
For each subject, we generate a random right censoring time $C_{i}\sim U(a,b)$ with $a=6$ and $b=22$. For the longitudinal process, we use $\Delta_{1}=1.6$, $\Delta_{2}=0.6$ for scenarios 1 and 3 and $\Delta_{1}=1.6$, $\Delta_{2}=2.2$ for scenario 2 to generate the visiting time points $t_{ij}$. 
For other parameters in the linear mixed model, we set $\beta_{1}=2$, $\beta_{2}=0.5$, and $\sigma_{e}=0.2$ for all scenarios. For scenario 3, the modified fixed effect parameters are set to $\beta_{1}^{'}=5$ and $\beta_{2}^{'}=-0.3$. 
These chosen simulation parameters characterize a marker exhibiting either a consistent upward trend over time (scenarios 1 and 2), or a structural change resulting in a reversed trend following the first transition (scenario 3). In all scenarios, higher marker levels are associated with a greater risk of all transitions. For model 2, we set $\Delta_{1}=2.6$, $\Delta_{2}=2$ and $\Delta_{3}=1.2$. Model parameter values are motivated by our application data. More specifically, to set the transition-specific parameters in the multistate submodel (i.e. $\{\gamma_{jk}\}$, $\{\alpha_{jk}\}$, $\{\delta_{jk}\}$ and $\{\lambda_{jk}\}$), we fit a separate joint model for each transition (as defined in Figure \ref{fig:mstate2}) using the JM-ST-C approach based on a random subset of subjects' data from the CPRD dataset. Parameters in the longitudinal submodel are obtained by fitting a linear mixed model (as specified in Section \ref{subsec:sim_models}) on the standardised log-transformed SBP based on subjects in the CPRD dataset who has had a diagnosis of T2D. From this we obtain $\beta_{1}=0.3$, $\beta_{2}=0.03$, $\sigma_{e}=0.6$ and $\sigma_{1}=0.7$, $\sigma_{2}=0.1$ and $\rho=-0.4$.

Motivated by \citet{alvares2021tractable}, we adopt the following weakly informative prior distributions in our posterior simulation for all the methods.
\begin{gather*} 
\alpha_{jk}, \gamma_{jk}, \beta_{1}. \beta_{2}\sim N(0, 100^2),\\ 
\lambda_{jk}, \delta_{jk}\sim \text{half-Cauchy}(0,1),\\
\sigma_{e}^{2}, \sigma_{1}^{2}, \sigma_{2}^{2}\sim \text{Inv-Gamma}(0.01,0.01),\\
\frac{\rho+1}{2}\sim \text{Be}(0.5,0.5),
\end{gather*}
where $\text{half-Cauchy}(\mu,\sigma)$ denotes a half-Cauchy distribution with location parameter $\mu$ and scale parameter $\sigma$, $\text{Inv-Gamma}(\alpha,\beta)$ denotes an inverse Gamma distribution with shape parameter $\alpha$ and scale parameter $\beta$ and $\text{Be}(\alpha,\beta)$ denotes a Beta distribution with shape parameters $\alpha$ and $\beta$. Of course, other reasonable priors could also be used here.

\subsection{Results} \label{subsec:sim_results}
We examined various sample sizes: $n=1000$ and $n=3000$ for Model 1 (for all three scenarios) and $n=1000$, $n=3000$ and $n=5000$ for Model 2.
In each case, we generated $N=100$ independent replications of the dataset. For the JM-CR and JM-ST approaches, parameters within each block were sampled from the respective posteriors in parallel, and the computation time was determined by taking the maximum computing time across all blocks. Posterior summaries were derived based on $1000$ NUTS samples, obtained using default control parameters for NUTS, after a suitable warm-up period $T_{b}$ that was selected based on preliminary runs to ensure convergence. We applied $T_{b}=300$ for $n=1000$ and $T_{b}=500$ for $n=3000$ and $n=5000$ for all approaches and simulation models. All computations were performed on the Cambridge Service for Data Driven Discovery (CSD3) High-Performance Computing (HPC) system using the Ice Lake CPUs. In this section, we report selected simulation results and briefly summarize the key findings. Additional details are provided in the appendix.

\begin{figure}[htbp]
  \centering
\includegraphics[width=1\textwidth]{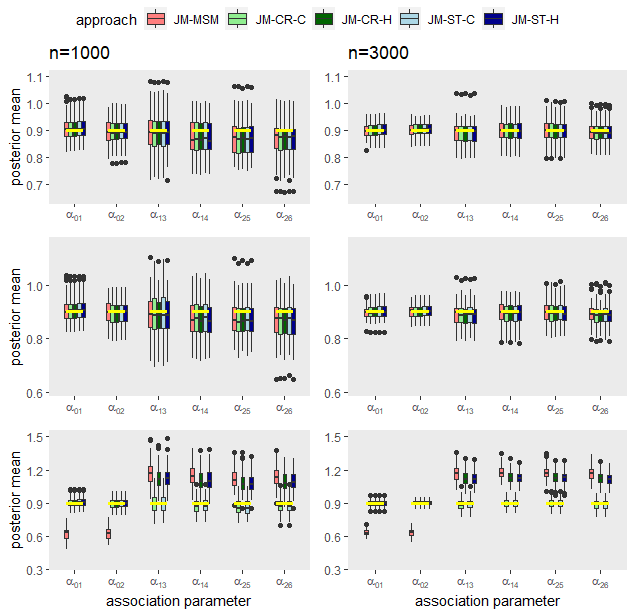}
 \caption{Box plot summary of the posterior mean of the association parameter $\alpha_{jk}$ obtained by each approach from $100$ replications of the data simulated from Model 1. Rows 1, 2 and 3 show the results for scenarios 1, 2 and 3, respectively. The left and right panels in each row are based on sample sizes of $n=1000$ and $n=3000$, respectively. The true parameter value is indicated by a yellow horizontal bar.}
  \label{fig:sim1results}
\end{figure}

\begin{figure}[htbp]
  \centering
\includegraphics[width=1\textwidth]{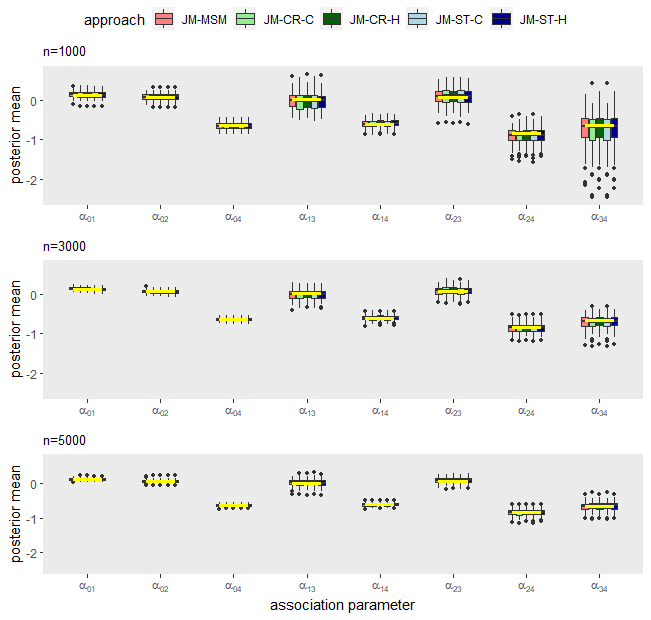}
 \caption{Box plot summary of the posterior mean of the association parameter $\alpha_{jk}$ obtained by each approach from $100$ replications of the data simulated from Model 2. 
 Rows 1, 2 and 3 show the results for $n=1000$, $n=3000$ and $n=5000$, respectively. Note that for $n=5000$, the result of the JM-MSM approach is based on 99 available repetitions as one replication exceeded the HPC computing limits and was terminated. The true parameter value is indicated by a yellow horizontal bar.}
  \label{fig:sim2results}
\end{figure}

In the joint model analysis, the association parameters $\alpha_{jk}$ are often of key interest. Figure \ref{fig:sim1results} summarizes the estimated posterior mean of the association parameter obtained by each approach across the 100 data replications for each scenario in Model 1. In scenarios 1 and 2 (upper and middle panels), the distributions of the point estimates obtained from the JM-CR and JM-ST approaches align with those from the JM-MSM approach. As the sample size $n$ increases, the variability of the estimates from all approaches reduces at a similar rate. In scenario 3 (lower panel), however, JM-MSM and blockwise approaches that utilize all historical longitudinal data (JM-CR-H and JM-ST-H) exhibit significant estimation biases, which persist with increasing sample sizes. This is unsurprising as the working models underlying these approaches are inherently misspecified: for JM-MSM, the multistate and longitudinal processes are not conditionally independent given the random effects, whereas for JM-CR-H and JM-ST-H, including additional longitudinal data would not improve estimation due to structural changes in the marker's dynamics. Still, blockwise approaches that use only concurrent longitudinal data (JM-CR-C and JM-ST-C) can yield accurate estimates in this scenario without needing to modify the model structure, and their accuracy improves as $n$ increases. Figure \ref{fig:sim2results} shows the point estimation results for $\alpha_{jk}$ for Model 2. All five approaches produce very consistent results for all sample sizes. The variability of the estimates tends to be much higher in later blocks compared to the first block, which could be attributed to a moderately large proportion of subjects being censored during the follow-up period.  As the sample size increases, this discrepancy diminishes for all approaches. We further compare posterior uncertainty quantification obtained from the five approaches. Tables \ref{tab:coverage_sim1} and \ref{tab:coverage_sim2} show the coverage probability (the proportion of times that the credible interval contains the true value of the parameter in repeated simulations) of the estimated $95\%$ credible intervals obtained from each approach. For scenarios 1 and 2 of Model 1, and for Model 2, all approaches showed similar coverage probabilities with slight variation around the theoretical value of $0.95$. In scenario 3 of Model 1, however, JM-MSM, JM-CR-H and JM-ST-H yield significantly lower coverage probabilities compared to JM-CR-C and JM-ST-C, and these probabilities decrease toward zero as $n$ increases. These results highlight the risk of model misspecification in the context of large data sets, where increasing certainty may lead to systematically incorrect results. Note that, although we focused on the association parameter here, the observed patterns also extend to other MSM parameters.

\begin{table}[htbp]
\centering
\caption{Coverage probability of estimated $95\%$ credible interval for the association parameter based on $100$ replications of data from simulation Model 1 ($n=1000$ \textbar{} $n=3000$).}
\label{tab:coverage_sim1}
\begin{tabular}{@{}ccccc@{}}
\toprule
Parameter & \multicolumn{1}{c}{Approach} & \multicolumn{1}{c}{Scenario 1} & \multicolumn{1}{c}{Scenario 2} & \multicolumn{1}{c}{Scenario 3} \\ \midrule
\multirow{5}{*}{$\alpha_{01}$} & JM-MSM  & 0.97 \textbar{} 0.97  & 0.96 \textbar{} 0.97  & 0.00 \textbar{} 0.00 \\
                            & JM-CR-C & 0.96 \textbar{} 0.95 & 0.97 \textbar{} 0.96 & 0.95 \textbar{} 0.95\\
                            & JM-CR-H & 0.96 \textbar{} 0.95 & 0.97 \textbar{} 0.96 & 0.95 \textbar{} 0.95\\
                            & JM-ST-C & 0.96 \textbar{} 0.95 & 0.97 \textbar{} 0.96& 0.97 \textbar{} 0.95\\
                            & JM-ST-H & 0.96 \textbar{} 0.95 & 0.97 \textbar{} 0.96& 0.97 \textbar{} 0.95\\\addlinespace
\multirow{5}{*}{$\alpha_{02}$} & JM-MSM  & 0.96 \textbar{} 0.96 & 0.94 \textbar{} 0.97& 0.00 \textbar{} 0.00\\
                            & JM-CR-C & 0.95 \textbar{} 0.97 & 0.95 \textbar{} 0.96& 0.93 \textbar{} 0.98\\
                            & JM-CR-H & 0.95 \textbar{} 0.97 & 0.95 \textbar{} 0.96& 0.93 \textbar{} 0.98\\
                            & JM-ST-C & 0.96 \textbar{} 0.97 & 0.95 \textbar{} 0.96& 0.93 \textbar{} 0.97\\
                            & JM-ST-H & 0.96 \textbar{} 0.97 & 0.95 \textbar{} 0.96&  0.93 \textbar{} 0.97\\\addlinespace
\multirow{5}{*}{$\alpha_{13}$} & JM-MSM  & 0.97 \textbar{} 0.93 & 0.96 \textbar{} 0.96& 0.20 \textbar{} 0.00\\
                            & JM-CR-C & 0.95 \textbar{} 0.95 & 0.93 \textbar{} 0.95& 0.93 \textbar{} 0.92\\
                            & JM-CR-H & 0.96 \textbar{} 0.93  & 0.95 \textbar{} 0.95& 0.32 \textbar{} 0.01\\
                            & JM-ST-C & 0.96 \textbar{} 0.94& 0.93 \textbar{} 0.94& 0.94 \textbar{} 0.93\\
                            & JM-ST-H & 0.96 \textbar{} 0.93& 0.94 \textbar{} 0.95& 0.27 \textbar{} 0.02\\\addlinespace
\multirow{5}{*}{$\alpha_{14}$} & JM-MSM  & 0.94 \textbar{} 0.97& 0.94 \textbar{} 0.94& 0.26 \textbar{} 0.00\\
                            & JM-CR-C & 0.94 \textbar{} 0.97& 0.92 \textbar{} 0.97& 0.96 \textbar{} 0.94\\
                            & JM-CR-H & 0.95 \textbar{} 0.97& 0.93 \textbar{} 0.93& 0.39 \textbar{} 0.00\\
                            & JM-ST-C & 0.94 \textbar{} 0.94& 0.94 \textbar{} 0.96& 0.96 \textbar{} 0.93\\
                            & JM-ST-H & 0.95 \textbar{} 0.96& 0.92 \textbar{} 0.94& 0.39 \textbar{} 0.00\\\addlinespace
\multirow{5}{*}{$\alpha_{25}$} & JM-MSM  & 0.95 \textbar{} 0.94& 0.95 \textbar{} 0.92& 0.35 \textbar{} 0.00\\
                            & JM-CR-C & 0.95 \textbar{} 0.91& 0.97 \textbar{} 0.92& 0.96 \textbar{} 0.92\\
                            & JM-CR-H & 0.96 \textbar{} 0.93& 0.96 \textbar{} 0.90& 0.56 \textbar{} 0.01\\
                            & JM-ST-C & 0.97 \textbar{} 0.91& 0.98 \textbar{} 0.91& 0.95 \textbar{} 0.92\\
                            & JM-ST-H & 0.94 \textbar{} 0.93& 0.95 \textbar{} 0.91& 0.54 \textbar{} 0.02\\\addlinespace
\multirow{5}{*}{$\alpha_{26}$} & JM-MSM  & 0.94 \textbar{} 0.94& 0.95 \textbar{} 0.94& 0.20 \textbar{} 0.00\\
                            & JM-CR-C & 0.94 \textbar{} 0.94& 0.92 \textbar{} 0.95& 0.98 \textbar{} 0.92\\
                            & JM-CR-H & 0.93 \textbar{} 0.92& 0.94 \textbar{} 0.94& 0.50 \textbar{} 0.01\\
                            & JM-ST-C & 0.94 \textbar{} 0.92& 0.93 \textbar{} 0.96& 0.96 \textbar{} 0.94\\
                            & JM-ST-H & 0.94 \textbar{} 0.95& 0.93 \textbar{} 0.94& 0.46 \textbar{} 0.04\\\bottomrule
\end{tabular}
\end{table}

\begin{table}[htbp]
\centering
\caption{Coverage probability of estimated $95\%$ credible interval for the association parameter based on $100$ replications of data from simulation Model 2 (Note: results for MSM with $n=5000$ are based on the $99$ available repetitions).}
\label{tab:coverage_sim2}
\begin{tabular}{@{}ccccc@{}}
\toprule
Parameter & \multicolumn{1}{c}{Approach} & \multicolumn{1}{c}{$n=1000$} & \multicolumn{1}{c}{$n=3000$} & \multicolumn{1}{c}{$n=5000$} \\ \midrule
\multirow{5}{*}{$\alpha_{01}$} & JM-MSM  & 0.95  & 0.98  & 0.98  \\
                              & JM-CR-C & 0.96 & 0.97 & 0.93 \\
                              & JM-CR-H & 0.96 & 0.97 & 0.93 \\
                              & JM-ST-C & 0.97 & 0.96 & 0.95 \\
                              & JM-ST-H & 0.97 & 0.96 & 0.95 \\\addlinespace
\multirow{5}{*}{$\alpha_{02}$} & JM-MSM  & 0.96 & 0.92 & 0.94 \\
                              & JM-CR-C & 0.96 & 0.95 & 0.93 \\
                              & JM-CR-H & 0.96 & 0.95 & 0.93 \\
                              & JM-ST-C & 0.96 & 0.95 & 0.95 \\
                              & JM-ST-H & 0.96 & 0.95 &  0.95 \\\addlinespace
\multirow{5}{*}{$\alpha_{04}$} & JM-MSM  & 0.94 & 0.96 & 0.92 \\
                             & JM-CR-C & 0.94 & 0.96 & 0.94 \\
                             & JM-CR-H & 0.94 & 0.96 & 0.94 \\
                             & JM-ST-C & 0.94 & 0.96 & 0.95 \\
                             & JM-ST-H & 0.94 & 0.96 & 0.95 \\\addlinespace
\multirow{5}{*}{$\alpha_{13}$} & JM-MSM  & 0.98 & 0.96 & 0.94 \\
                             & JM-CR-C & 0.95 & 0.97 & 0.96 \\
                             & JM-CR-H & 0.98 & 0.96 & 0.94 \\
                             & JM-ST-C & 0.96 & 0.96 & 0.95 \\
                             & JM-ST-H & 0.98 & 0.95 & 0.95 \\\addlinespace
\multirow{5}{*}{$\alpha_{14}$} & JM-MSM  & 0.97 & 0.93 & 0.95 \\
                             & JM-CR-C & 0.97 & 0.93 & 0.96 \\
                             & JM-CR-H & 0.96 & 0.91 & 0.95 \\
                             & JM-ST-C & 0.97 & 0.93 & 0.96 \\
                             & JM-ST-H & 0.96 & 0.93 & 0.95 \\\addlinespace
\multirow{5}{*}{$\alpha_{23}$} & JM-MSM  & 0.95 & 0.95 & 0.95 \\
                             & JM-CR-C & 0.95 & 0.95 & 0.96 \\
                             & JM-CR-H & 0.95 & 0.93 & 0.96 \\
                             & JM-ST-C & 0.96 & 0.94 & 0.96 \\
                             & JM-ST-H & 0.95 & 0.93 & 0.97 \\\addlinespace
\multirow{5}{*}{$\alpha_{24}$} & JM-MSM  & 0.95 & 0.95 & 0.95 \\
                             & JM-CR-C & 0.95 & 0.95 & 0.96 \\
                             & JM-CR-H & 0.93 & 0.95 & 0.95 \\
                             & JM-ST-C & 0.96 & 0.94 & 0.97 \\
                             & JM-ST-H & 0.94 & 0.95 & 0.95 \\\addlinespace                            
\multirow{5}{*}{$\alpha_{34}$} & JM-MSM  & 0.92 & 0.96 & 0.93 \\
                             & JM-CR-C & 0.93 & 0.95 & 0.95 \\
                             & JM-CR-H & 0.92 & 0.97 & 0.93 \\
                             & JM-ST-C & 0.93 & 0.95 & 0.95 \\
                             & JM-ST-H & 0.92 & 0.97 & 0.93\\\bottomrule
\end{tabular}
\end{table}

\begin{figure}[htbp]
  \centering
  \includegraphics[width=0.95\textwidth]{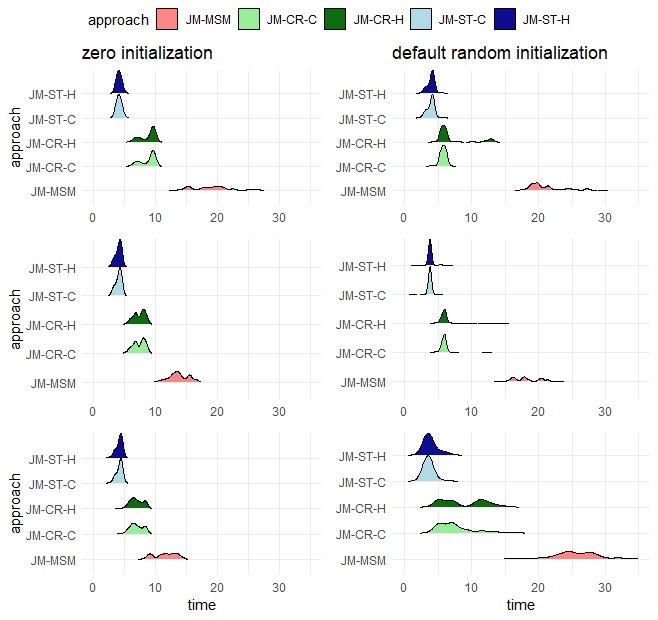}
  \caption{Distribution of computation times required for obtaining $1000$ convergent samples, including burn-in, for each approach across $100$ replications of the data ($n=3000$) for Model 1. Kernel density estimation was used to obtain the density plots, with the default bandwidth selected by the ggridges R package. The left panel shows computing times when all parameters are initialized at 0 on the transformed unconstrained space, while the right panel shows computing times when all parameters are randomly initialized over the interval $[-2,2]$ on the transformed unconstrained space (default in Rstan).
  }
  \label{fig:sim1time}
\end{figure}

\begin{figure}[htbp]
  \centering
  \includegraphics[width=0.95\textwidth]{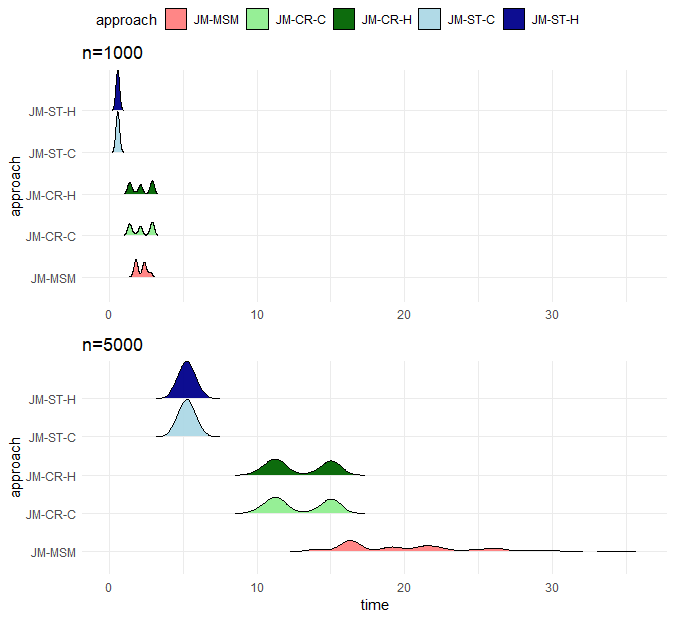}
  \caption{Distribution of computation times required for obtaining $1000$ convergent samples, including burn-in, for each approach across $100$ replications of the data for Model 2. Kernel density estimation was used to obtain the density plots, with the default bandwidth selected by the ggridges R package. The upper panel corresponds to $n=1000$, while the lower panel corresponds to $n=5000$ (for JM-MSM approach, the plot is based on 99 available repetitions). In both cases, all parameters were initialized at 0 on the transformed unconstrained space.
  }
  \label{fig:sim2time}
\end{figure}

Figures \ref{fig:sim1time} and \ref{fig:sim2time} display selected summary results of the computation time (including the pre-specified warm-up period) needed to obtain 1000 MCMC samples after convergence for Models 1 and 2, respectively. As expected, the JM-ST approach is the most efficient in all scenarios, while the JM-MSM approach requires the longest run time. For the blockwise approaches, the two sub-versions, which differ only in the construction of the longitudinal submodel, have very similar computational costs. We also  found that the initialization strategy of the NUTS and the sample size have a more substantial impact on the JM-MSM approach compared to the blockwise approaches (as seen when comparing the left and right panels of Figure \ref{fig:sim1time} and the top and bottom panels of Figure \ref{fig:sim2time}). With random/overdispersed initialization or increasing sample sizes, JM-MSM typically requires a longer warm-up period for reaching convergence, with JM-ST being the least influenced. Therefore, as the sample size grows, we expect the computational advantage of the blockwise approach over JM-MSM to become more prominent.

We also evaluated the feasibility of the LOO-CV for guiding the selection of the strategy for linking longitudinal data with model blocks when implementing the blockwise approaches, i.e., choosing between JM-CR-C/JM-ST-C or JM-CR-H/JM-ST-H for a specific block (note that this issue is not relevant for block(s) involving state 0, as there is no history to consider). 
Our findings indicate that LOO-CV computed based on data definitions of i) and iii) (see Section \ref{subsec:modelcompare}) yield satisfactory results, in the sense that they successfully identify the approach that better reflects the data-generating process.
For scenarios 1 and 2 of Model 1 and for Model 2, the approach using all historical longitudinal data is consistently preferred over the alternative option for each block in all repetitions of the data.
For scenario 3 of Model 1, the approach using only longitudinal data within the block is strongly favoured over the other option, which aligns with our expectations due to the presence of a change point in the longitudinal trajectory.
It is interesting to note that for this specific model comparison problem, LOO-CV computed based solely on time-to-event data lead to ``inaccurate" results in some cases. This may not be too surprising, considering that the sole difference between the two configurations lies in the longitudinal submodel, and thus the predictive accuracy of the longitudinal data becomes more relevant in this context. Focusing solely on the predictive accuracy of the time-to-event data could be misleading as the parameters in the survival submodel undergo adjustments (for the estimated longitudinal trajectory) during the fitting process.

\section{Application to the CPRD data} \label{sec:appl}
Multimorbidity, defined as the co-existence of two or more chronic conditions in an individual, is becoming increasingly prevalent and poses significant challenges to individuals and public health \citep{wallace2015managing}.
There is an important need to better understand the longitudinal accumulation of diseases for improved management of multimorbidity \citep{cezard2021studying}. 
Here we illustrate the use of the proposed approaches by analysing the association between routinely measured SBP and the progression of multimorbidity defined as the combinations of three common chronic conditions, namely type-2 diabetes (T2D), mental health conditions (MH) and cardiovascular diseases (CVD), based on the CPRD Aurum database. The CPRD Aurum contains information on individual patient demographics, clinical observations, diagnoses, and treatments, and we refer to \citet{wolf2019data} for more background information.
Here, for our analysis, we modelled the standardized log-transformed SBP using a linear mixed model, incorporating a random intercept and slope as in Equation \eqref{eq:sim_LMM}. The disease progression was modelled using the multistate process depicted in Figure \ref{fig:mstate2}, where the transition intensities were specified as in Equation \eqref{eq:sim_intensity_MSM}, and the standardized age at entry to the current state was considered as a prognostic covariate (i.e. $w_{i}$).

\begin{figure}
  \centering
  \includegraphics[width=0.95\textwidth]{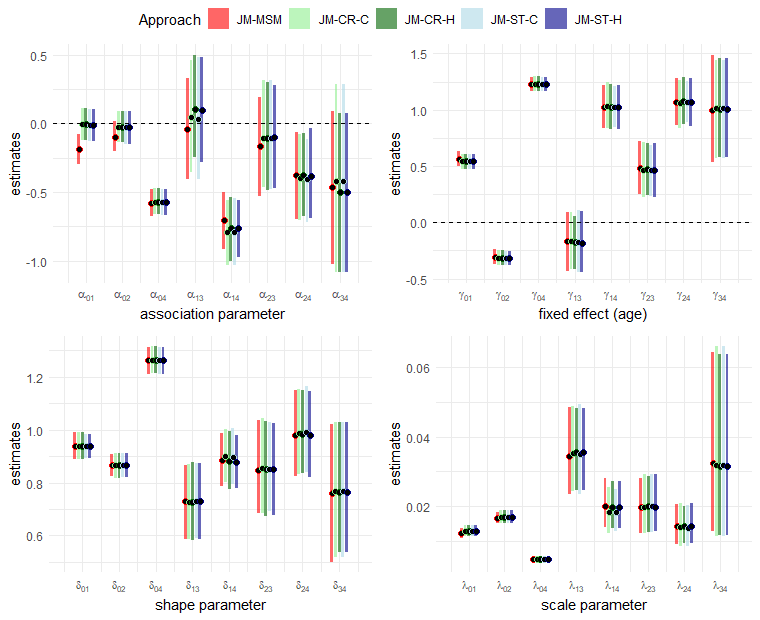}
  \caption{Estimation results obtained by the JM-MSM, JM-CR-C, JM-CR-H, JM-ST-C and JM-ST-H approaches. The black dot represents the posterior mean and the rectangular bar superimposed represents the corresponding $95\%$ credible interval. 
  }
  \label{fig:app_5methodscompare}
\end{figure}

First, we compared the five estimation approaches - JM-MSM, JM-CR-H, JM-CR-C, JM-ST-H, and JM-ST-C, as considered in the simulation study.
Our analysis dataset was derived from the multimorbidity cohort as extracted and detailed in \citet{chen2023effect}, by focusing on the subjects with an initial T2D diagnosis (among the 3 conditions) and had at least one SBP measurement recorded during their follow-up period (any SBP measurements taken within 3 months before death were removed to reduce the potential confounding effects of the near-death period on the SBP). To allow for proper implementation and comparison of the approaches, we employed a popular imputation strategy to handle the issue of missing longitudinal data within a block (relevant for the blockwise approaches).
In situations where a subject had no recorded SBP measurement within block $B_{1}$ (i.e. from the T2D diagnosis time until the first transition), we utilized the most recent SBP measurement recorded prior to the T2D diagnosis as an imputed baseline SBP.
For subsequent blocks (apart from $B_{1}$), when using blockwise approaches based on concurrent longitudinal data, JM-CR-C or JM-ST-C, for inference, we applied the same imputation strategy to assign an SBP value at the time of entering into the block if a subject lacks an SBP reading within the block. 
Here for our comparison, we used a randomly selected and processed subset of the cohort, comprised of $n=15142$ subjects. This sample size is about the largest manageable size for JM-MSM under the computational constraints of the HPC. 
The posterior inference was based on 1000 NUTS samples after a warm-up period of $T_{b}=700$ for JM-MSM and $T_{b}=500$ for JM-CR and JM-ST approaches. Using the previously specified computing resources, the entire sampling process took approximately $35.2$, $30$ and $8.7$ hours for JM-MSM, JM-CR-C/JM-CR-H, and JM-ST-C/JM-ST-H, respectively. As expected, JM-MSM was the most computationally expensive while JM-ST was the most efficient. Note that JM-CR provided a very limited efficiency gain over JM-MSM. This is largely attributed to the fact that for this dataset, a significant proportion of subjects are censored within block 1, which dominates the computing burden.
Figure \ref{fig:app_5methodscompare} displays the posterior mean and the associated $95\%$ credible intervals for both the association and other MSM parameters, obtained by each of the five approaches. 
The estimated fixed effect and Weibull parameters generally align across the approaches. However, minor discrepancies are observed for the estimated association parameter for several transitions, particularly between JM-MSM, JM-CR-C/JM-ST-C, and JM-CR-H/JM-ST-H.
These differences could be indicative of potential changes in the longitudinal trajectory following disease progression, such that different subsets of the longitudinal data lead to different estimated longitudinal trajectories, which in turn influence the association parameter. In such scenarios, blockwise approaches JM-CR-C/JM-ST-C are expected to be more adaptive and robust to longitudinal model misspecification.
For further evidence, we compare JM-ST-C and JM-ST-H for each transition using the LOO-CV as described in Section \ref{subsec:modelcompare}. 
The results indicate a consistent preference for JM-ST-C across all transitions, suggesting the potential change points in the longitudinal trajectory.

Motivated by the comparison results above, we proceed with the most efficient JM-ST-C approach, which enables us to leverage a more substantial amount of data for estimating parameters associated with each transition.
Table \ref{tab:app_JM-ST_datacounts} shows the number of subjects' data that were used for estimating each transition, noting that we utilised all available data in our cohort for estimating transitions $1\to 3$,  $1\to 4$,  $2\to 3$,  $2\to 4$ and $3\to 4$. In total, this corresponds to $n=72735$ subjects' data being used for model fitting.
For each transition, we explored three different association structures.
Model 1 (M1) includes only the current underlying value of the longitudinal
marker SBP. Model 2 (M2) includes the current value of SBP and an interaction effect between age and SBP. Model 3 (M3) extends M1 by adding a quadratic association term, thus allowing for a non-linear relationship between the current value of SBP and the rate of transition.
The transition intensities under M1, M2 and M3 are therefore given by
\begin{align*}
&\text{M1}: h_{jk}^{(i)}(B(t))=h_{0,jk}(B(t))\exp(\text{age}_{i}\gamma_{jk}+\alpha_{jk}\mu_{i}(t)) \\[1em]
&\text{M2}: h_{jk}^{(i)}(B(t))=h_{0,jk}(B(t))\exp(\text{age}_{i}\gamma_{jk}+\alpha_{jk,1}\mu_{i}(t)+\alpha_{jk,2}(\text{age}_{i}\times \mu_{i}(t))) \\[1em]
&\text{M3}: h_{jk}^{(i)}(B(t))=h_{0,jk}(B(t))\exp(\text{age}_{i}\gamma_{jk}+\alpha_{jk,1}\mu_{i}(t)+\alpha_{jk,2}\mu_{i}^{2}(t))
\end{align*}

We implemented the JM-CR-C approach under each model, where inference was based on 1000 NUTS samples following a burn-in of 500 samples (convergence diagnostics suggest that this is sufficient).
The total sampling time required for M1, M2 and M3 is approximately 29.6, 34.2 and 31.8 hours, respectively. 
We compared M1, M2 and M3 for each transition using the LOO-CV computed based on both longitudinal and time-to-event data (version iii as described in Section \ref{subsec:modelcompare}). 
Table \ref{tab:app_JMSTresults} shows the posterior summaries of the association parameter(s) obtained under each model, with the favoured model for each transition highlighted in blue shade.
It is noteworthy that different association structures were suggested for different pathways of disease transition. 
Specifically, for transitions leading to death ($0\to 4$, $1\to 4$, $2\to 4$ and $3\to 4$), a quadratic association pattern is suggested.
The coefficient of the quadratic effect, $\alpha_{ij,2}$, maintains a positive value across all transitions, thus indicating that both lower or higher SBP values are linked with an increased rate of mortality. The precise quadratic relationship (controlled by both $\alpha_{ij,1}$ and $\alpha_{ij,2}$) varies depending on the subject's comorbidity status prior to death.
While it is well established that high SBP is a risk factor for mortality, 
our finding of an elevated risk at lower SBP levels is less commonly recognized. However, our results aligh with recent reports based on the Cox regression analysis \citep{vamos2012association,masoli2020blood}.
Furthermore, the association pattern of SBP level on the rate of transitioning into CVD ($0\to 1$ and $2\to 3$) or MH ($0\to 2$ and $1\to 3$) is found to be influenced by the underlying comorbidity status.
For instance, for transitions into CVD, a quadratic association with SBP is identified when T2D is the only existing condition. However, when MH is present as an additional existing comorbidity, an interaction effect between SBP and age is suggested.
This highlights the complexity of the relationship between a health marker like SBP and the temporal trajectory of multimorbidity.

\begin{table}[ht!]
\centering
\caption{Number of subjects' data used for estimating each disease transition using the JM-ST-C approach.}
\label{tab:app_JM-ST_datacounts}
\begin{tabular}{lcccccc}
\toprule
\textbf{Transition} & $0\to 1$, $0\to 2$, $0\to 4$ & $1\to 3$ & $1\to 4$ & $2\to 3$ & $2\to 4$ & $3\to 4$\\
\midrule
\textbf{Sample size} & 37889  & 20018 & 20017 & 19903 & 19901 & 2647\\
\bottomrule
\end{tabular}
\end{table}

\begin{table}[htbp]
\centering
\caption{Posterior summary of the association parameter estimated using JM-ST-C approach under models M1, M2, and M3. The favoured model based on the LOO-CV for each transition is highlighted in blue shade.}
\label{tab:app_JMSTresults}
\begin{tabular}{lcccc}
\toprule
\multicolumn{5}{c}{M1} \\
\midrule
Disease Transition & Parameter & Mean & 2.5\% & 97.5\% \\
\midrule
T2D $\to$ T2D+CVD & $\alpha_{01}$ & 0.010 & -0.061 & 0.080 \\
T2D $\to$ T2D+MH & $\alpha_{02}$ & -0.087 & -0.160 & -0.020 \\
T2D $\to$ Death & $\alpha_{04}$ & -0.624 & -0.682 & -0.566 \\
\rowcolor{highlight} T2D+CVD $\to$ T2D+CVD+MH & $\alpha_{13}$ & \textbf{0.068} & \textbf{-0.038} & \textbf{0.172} \\
T2D+CVD $\to$ Death & $\alpha_{14}$ & -0.572 & -0.623 & -0.519 \\
T2D+MH $\to$ T2D+MH+CVD & $\alpha_{23}$ & 0.099 & -0.002 & 0.198 \\
T2D+MH $\to$ Death & $\alpha_{24}$ & -0.696 & -0.773 & -0.617 \\
T2D+MH+CVD $\to$ Death & $\alpha_{34}$ & -0.640 & -0.792 & -0.487 \\
\midrule
\multicolumn{5}{c}{M2} \\
\midrule
Disease Transition & Parameter & Mean & 2.5\% & 97.5\% \\
\midrule
T2D $\to$ T2D+CVD & $\alpha_{01,1}$  & 0.077 & -0.006 & 0.164 \\
         & $\alpha_{01,2}$ & -0.123 & -0.200 & -0.051 \\
\rowcolor{highlight} T2D $\to$ T2D+MH & \textbf{$\alpha_{02,1}$} & \textbf{-0.078} & \textbf{-0.151} & \textbf{-0.003} \\
\rowcolor{highlight}         & \textbf{$\alpha_{02,2}$} & \textbf{0.024} & \textbf{-0.044} & \textbf{0.091} \\
T2D $\to$ Death & $\alpha_{04,1}$ & -0.518 & -0.619 & -0.417 \\
         & $\alpha_{04,2}$ & -0.097 & -0.166 & -0.025 \\
T2D+CVD $\to$ T2D+CVD+MH & $\alpha_{13,1}$ & 0.069  & -0.040  & 0.183  \\
         & $\alpha_{13,2}$ & 0.009  & -0.095 & 0.113 \\
T2D+CVD $\to$ Death & $\alpha_{14,1}$ & -0.608 & -0.674 & -0.541 \\
         & $\alpha_{14,2}$ & 0.054 & -0.017 &  0.120 \\
\rowcolor{highlight} T2D+MH $\to$ T2D+MH+CVD & \textbf{$\alpha_{23,1}$} & \textbf{0.147} & \textbf{0.037} &  \textbf{0.256} \\
\rowcolor{highlight}         & \textbf{$\alpha_{23,2}$} & \textbf{-0.127} & \textbf{-0.227} & \textbf{-0.023} \\
T2D+MH $\to$ Death & $\alpha_{24,1}$ & -0.553 & -0.662 & -0.435 \\
         & $\alpha_{24,2}$ & -0.138 & -0.231 & -0.053 \\
T2D+MH+CVD $\to$ Death & $\alpha_{34,1}$ & -0.503 & -0.701 &  -0.305 \\
         & $\alpha_{34,2}$ & -0.216 &  -0.392 & -0.034 \\
\midrule
\multicolumn{5}{c}{M3} \\
\midrule
Disease Transition & Parameter & Mean & 2.5\% & 97.5\% \\
\midrule
\rowcolor{highlight} T2D $\to$ T2D+CVD & \textbf{$\alpha_{01,1}$}  & \textbf{0.089} & \textbf{0.021} & \textbf{0.160} \\
\rowcolor{highlight}         & \textbf{$\alpha_{01,2}$} & \textbf{0.194} & \textbf{0.135} & \textbf{0.250} \\
T2D $\to$ T2D+MH & $\alpha_{02,1}$ & -0.027 & -0.109 & 0.049 \\
         & $\alpha_{02,2}$ & 0.096 & 0.028 & 0.162 \\
\rowcolor{highlight} T2D $\to$ Death & \textbf{$\alpha_{04,1}$} & \textbf{-0.373} & \textbf{-0.444} & \textbf{-0.307} \\
\rowcolor{highlight}         & \textbf{$\alpha_{04,2}$} & \textbf{0.263} & \textbf{0.215} & \textbf{0.313} \\
T2D+CVD $\to$ T2D+CVD+MH & $\alpha_{13,1}$ & 0.112  & -0.008  & 0.233  \\
         & $\alpha_{13,2}$ & 0.069  & -0.044 & 0.166 \\
\rowcolor{highlight} T2D+CVD $\to$ Death & \textbf{$\alpha_{14,1}$} & \textbf{-0.300} & \textbf{-0.375} & \textbf{-0.228} \\
\rowcolor{highlight}          & \textbf{$\alpha_{14,2}$} & \textbf{0.219} & \textbf{0.166} &  \textbf{0.268} \\
T2D+MH $\to$ T2D+MH+CVD & $\alpha_{23,1}$ & 0.170 & 0.080 &  0.255 \\
         & $\alpha_{23,2}$ & 0.250 & 0.173 & 0.326 \\
\rowcolor{highlight} T2D+MH $\to$ Death & \textbf{$\alpha_{24,1}$} & \textbf{-0.422} & \textbf{-0.510} & \textbf{-0.345} \\
 \rowcolor{highlight}         & \textbf{$\alpha_{24,2}$} & \textbf{0.312} & \textbf{0.246} & \textbf{0.378} \\
\rowcolor{highlight} T2D+MH+CVD $\to$ Death & \textbf{$\alpha_{34,1}$} & \textbf{-0.374} & \textbf{-0.601} &  \textbf{-0.164} \\
 \rowcolor{highlight}        & \textbf{$\alpha_{34,2}$} & \textbf{0.250} & \textbf{0.090} & \textbf{0.399} \\
\bottomrule
\end{tabular}
\end{table}

\section{Discussion} \label{sec:disc}
In this paper, we proposed blockwise approaches for inference in joint longitudinal and multistate models for the first time, exploiting parallel computing and state-of-the-art sampling techniques. 
Our simulation study demonstrates that these approaches offer notable computational efficiency gains over the standard estimation strategy while maintaining accurate estimation and allowing straightforward implementation. Model selection/comparison can be performed efficiently in a blockwise manner using the Bayesian leave-one-out cross-validation criterion, LOO-CV. 
The practical feasibility and scalability of our proposed methods are demonstrated through an application to real-world data. We analysed the concurrent evolution of SBP and the progression of comorbidity arising from T2D, CVD and MH, based on a large UK electronic health record dataset. Our analysis revealed distinct association structures between SBP levels and different disease accumulation pathways. 
It should be noted that, although our focus has been on the case of a single longitudinal marker and two specific multistate processes under the semi-Markov assumption, our proposed approaches can be easily extended to broader scenarios. For instance, they can accommodate multivariate longitudinal biomarkers and more general and complex multistate processes (e.g. non-Markov processes) by appropriately defining the blocks and regression models for the longitudinal and multistate processes. We anticipate that the advantages of using these blockwise approaches will be even more prominent in these more complex scenarios. Furthermore, we believe that there is potential to exploit another level of parallelism during the sampling stage, i.e. by running multiple independent chains in parallel, each generating a fraction of the total desired samples, and then combining samples post-convergence. Given a fixed burn-in period and a set total number of samples post-combination, we expect the blockwise approaches to reap more efficiency gains than JM-MSM, as the burn-in period often constitutes a much larger proportion of the total sampling time for the latter than the former.
The recent two-stage approach proposed in \citet{alvares2023} for joint model estimation can also be employed alongside our blockwise approaches to further accelerate the inference in each block. The core idea involves estimating parameters related to each block in two stages: each stage estimates a part of the full parameter set, thereby reducing the overall computational load compared to a single-stage joint estimation process. Under certain conditions, this procedure has been shown to asymptotically resemble a joint inference.

The methodology we have developed here opens several venues for future research. An important downstream application of the joint modelling framework is individual dynamic prediction, which involves obtaining dynamically updated, subject-specific predictions of event risk using all available information up to the time of prediction. While this task is well-studied in the context of joint models involving survival or competing risk event processes \citep{rizopoulos2011dynamic,ferrer2019individual}, it is much less explored in the presence of multistate data \citep{ferrer2016joint}. Our proposed blockwise approach offers new strategies for efficiently performing certain dynamic prediction tasks and assessing associated uncertainty under this latter setting. For instance, with our joint competing risk approach, estimators proposed for competing risk settings, see e.g. \citet{ferrer2019individual}, could be adapted and employed for estimating transition/state dwell probabilities within a competing risk block based on our block-specific posteriors. However, for more general types of transition probabilities (e.g. among states across different blocks), efficient predictive inference remains a challenging task, due to the fact these quantities typically do not admit explicit expressions in terms of the model parameters. 
In addition, our current modelling framework presumes that the multistate process is continuously observed, i.e. the exact time of transition is known. 
However, in practical applications, these transition times may, by their nature, be subject to interval censoring, which would bring additional computational challenges due to the intractability of the likelihood function for the resulting multistate data \citep{lovblom2023joint}. It would be of interest to explore efficient estimation strategies in these settings.
Lastly, approximate Bayesian inference methods such as variational inference (VI) or Laplace-based methods offer a potential alternative to our sampling-based approaches. These methods are often expected to be computationally cheaper due to their deterministic nature. In our simulation study, we experimented with the ``vanilla'' VI method implemented in {\tt rstan}, based on either mean-field or Gaussian full-rank approximations of the posterior. We found however that the algorithm suffered from serious convergence issues in all scenarios. We leave further investigation and comparison to more advanced VI or related approximate algorithms as future work.

\bibliographystyle{apalike}
\bibliography{sample}

\begin{thebibliography}{}

\bibitem[Alvares and Leiva-Yamaguchi, 2023]{alvares2023}
Alvares, D. and Leiva-Yamaguchi, V. (2023).
\newblock A two-stage approach for {B}ayesian joint models: {R}educing
  complexity while maintaining accuracy.
\newblock {\em Statistics and Computing}, 33(5):1--11.

\bibitem[Alvares and Rubio, 2021]{alvares2021tractable}
Alvares, D. and Rubio, F.~J. (2021).
\newblock A tractable {B}ayesian joint model for longitudinal and survival
  data.
\newblock {\em Statistics in Medicine}, 40(19):4213--4229.

\bibitem[Andrinopoulou and Rizopoulos, 2016]{andrinopoulou2016bayesian}
Andrinopoulou, E.~R. and Rizopoulos, D. (2016).
\newblock Bayesian shrinkage approach for a joint model of longitudinal and
  survival outcomes assuming different association structures.
\newblock {\em Statistics in Medicine}, 35(26):4813--4823.

\bibitem[Betancourt, 2017]{betancourt2017conceptual}
Betancourt, M. (2017).
\newblock A conceptual introduction to {H}amiltonian {M}onte {C}arlo.
\newblock {\em arXiv preprint arXiv:1701.02434}.

\bibitem[Cezard et~al., 2021]{cezard2021studying}
Cezard, G., {McHale}, C.~T., Sullivan, F., Bowles, J. K.~F., and Keenan, K.
  (2021).
\newblock Studying trajectories of multimorbidity: {A} systematic scoping
  review of longitudinal approaches and evidence.
\newblock {\em BMJ Open}, 11:1--19.

\bibitem[Chen et~al., 2023]{chen2023effect}
Chen, S., Marshall, T., Jackson, C., Cooper, J., Crowe, F., Nirantharakumar,
  K., Saunders, C.~L., Kirk, P., Richardson, S., Edwards, D., Griffin, S., Yau,
  C., and Barrett, J.~K. (2023).
\newblock Effect of sociodemographic characteristics on longitudinal
  progression of multimorbidity: {A} multistate modelling analysis of a large
  primary care records dataset.
\newblock {\em medRxiv preprint 10.1101/2023.03.06.23286491v1}.

\bibitem[Cook and Lawless, 2018]{cook2018multistate}
Cook, R.~J. and Lawless, J.~F. (2018).
\newblock {\em Multistate models for the analysis of life history data}.
\newblock Chapman \& Hall/CRC.

\bibitem[Cox and Hinkley, 1979]{cox1979theoretical}
Cox, D.~R. and Hinkley, D.~V. (1979).
\newblock {\em Theoretical statistics}.
\newblock CRC Press.

\bibitem[Crowther and Lambert, 2017]{crowther2017parametric}
Crowther, M.~J. and Lambert, P.~C. (2017).
\newblock Parametric multistate survival models: {F}lexible modelling allowing
  transition-specific distributions with application to estimating clinically
  useful measures of effect differences.
\newblock {\em Statistics in Medicine}, 36(29):4719--4742.

\bibitem[Dantan et~al., 2011]{dantan2011joint}
Dantan, E., Joly, P., Dartigues, J.~F., and {Jacqmin-Gadda}, H. (2011).
\newblock Joint model with latent state for longitudinal and multistate data.
\newblock {\em Biostatistics}, 12(4):723--736.

\bibitem[Dessie et~al., 2020]{dessie2020modelling}
Dessie, Z.~G., Zewotir, T., Mwambi, H., and North, D. (2020).
\newblock Modelling of viral load dynamics and {CD4} cell count progression in
  an antiretroviral naive cohort: {U}sing a joint linear mixed and multistate
  {M}arkov model.
\newblock {\em BMC Infectious Diseases}, 20:1--14.

\bibitem[Ferrer et~al., 2019]{ferrer2019individual}
Ferrer, L., Putter, H., and {Proust-Lima}, C. (2019).
\newblock Individual dynamic predictions using landmarking and joint modelling:
  {V}alidation of estimators and robustness assessment.
\newblock {\em Statistical Methods in Medical Research}, 28(12):3649--3666.

\bibitem[Ferrer et~al., 2016]{ferrer2016joint}
Ferrer, L., Rondeau, V., Dignam, J., Pickles, T., {Jacqmin-Gadda}, H., and
  {Proust-Lima}, C. (2016).
\newblock Joint modelling of longitudinal and multi-state processes:
  {A}pplication to clinical progressions in prostate cancer.
\newblock {\em Statistics in Medicine}, 35(22):3933--3948.

\bibitem[Freisling et~al., 2020]{freisling2020lifestyle}
Freisling, H., Viallon, V., Lennon, H., Bagnardi, V., Ricci, C., Butterworth,
  A.~S., Sweeting, M., Muller, D., Romieu, I., Bazelle, P., et~al. (2020).
\newblock Lifestyle factors and risk of multimorbidity of cancer and
  cardiometabolic diseases: {A} multinational cohort study.
\newblock {\em BMC Medicine}, 18:1--11.

\bibitem[Furgal, 2021]{furgal2021bayesian}
Furgal, A. (2021).
\newblock {\em Bayesian models for joint longitudinal and multi-state survival
  data}.
\newblock PhD thesis, University of Michigan.

\bibitem[Gelman et~al., 2013]{gelman2013bayesian}
Gelman, A., Carlin, J.~B., Stern, H.~S., Dunson, D.~B., Vehtari, A., and Rubin,
  D.~B. (2013).
\newblock {\em Bayesian data analysis}.
\newblock CRC press.

\bibitem[Hickey et~al., 2016]{hickey2016joint}
Hickey, G.~L., Philipson, P., Jorgensen, A., and {Kolamunnage-Dona}, R. (2016).
\newblock Joint modelling of time-to-event and multivariate longitudinal
  outcomes: {R}ecent developments and issues.
\newblock {\em BMC Medical Research Methodology}, 16:1--15.

\bibitem[Hickey et~al., 2018a]{hickey2018comparison}
Hickey, G.~L., Philipson, P., Jorgensen, A., and {Kolamunnage-Dona}, R.
  (2018a).
\newblock A comparison of joint models for longitudinal and competing risks
  data, with application to an epilepsy drug randomized controlled trial.
\newblock {\em Journal of the Royal Statistical Society. Series A (Statistics
  in Society)}, 181(4):1105--1123.

\bibitem[Hickey et~al., 2018b]{hickey2018joint}
Hickey, G.~L., Philipson, P., Jorgensen, A., and {Kolamunnage-Dona}, R.
  (2018b).
\newblock Joint models of longitudinal and time-to-event data with more than
  one event time outcome: {A} review.
\newblock {\em The International Journal of Biostatistics}, 14(1):1--19.

\bibitem[Hoffman and Gelman, 2014]{hoffman2014no}
Hoffman, M.~D. and Gelman, A. (2014).
\newblock The {No-U-Turn} sampler: {A}daptively setting path lengths in
  {H}amiltonian {M}onte {C}arlo.
\newblock {\em Journal of Machine Learning Research}, 15(1):1593--1623.

\bibitem[Ibrahim et~al., 2010]{ibrahim2010basic}
Ibrahim, J.~G., Chu, H., and Chen, L.~M. (2010).
\newblock Basic concepts and methods for joint models of longitudinal and
  survival data.
\newblock {\em Journal of Clinical Oncology}, 28(16):2796--2801.

\bibitem[Lovblom, 2023]{lovblom2023joint}
Lovblom, L.~E. (2023).
\newblock {\em Joint Multistate Models for Correlated Disease Processes:
  Extending Approaches for Interval-Censoring, Mixed Observation Schemes, and
  Multiple Longitudinal Outcomes}.
\newblock PhD thesis, University of Toronto (Canada).

\bibitem[Masoli et~al., 2020]{masoli2020blood}
Masoli, J. A.~H., Delgado, J., Pilling, L., Strain, D., and Melzer, D. (2020).
\newblock Blood pressure in frail older adults: {A}ssociations with
  cardiovascular outcomes and all-cause mortality.
\newblock {\em Age and Ageing}, 49(5):807--813.

\bibitem[Neal, 2011]{neal2011mcmc}
Neal, R.~M. (2011).
\newblock {\em Handbook of {M}arkov chain {M}onte {C}arlo}, chapter {MCMC}
  using {H}amiltonian dynamics, pages 113--162.
\newblock Chapman \& Hall/CRC.

\bibitem[Papageorgiou et~al., 2019]{papageorgiou2019overview}
Papageorgiou, G., Mauff, K., Tomer, A., and Rizopoulos, D. (2019).
\newblock An overview of joint modeling of time-to-event and longitudinal
  outcomes.
\newblock {\em Annual Review of Statistics and Its Application}, 6:223--240.

\bibitem[Rizopoulos, 2011]{rizopoulos2011dynamic}
Rizopoulos, D. (2011).
\newblock Dynamic predictions and prospective accuracy in joint models for
  longitudinal and time-to-event data.
\newblock {\em Biometrics}, 67(3):819--829.

\bibitem[Rizopoulos, 2012]{rizopoulos2012joint}
Rizopoulos, D. (2012).
\newblock {\em Joint models for longitudinal and time-to-event data: {W}ith
  applications in {R}}.
\newblock Chapman \& Hall/CRC.

\bibitem[Robert et~al., 1999]{robert1999monte}
Robert, C.~P., Casella, G., and Casella, G. (1999).
\newblock {\em Monte Carlo statistical methods}, volume~2.
\newblock Springer.

\bibitem[{Singh-Manoux} et~al., 2018]{singh2018clinical}
{Singh-Manoux}, A., Fayosse, A., Sabia, S., Tabak, A., Shipley, M., Dugravot,
  A., and Kivim{\"a}ki, M. (2018).
\newblock Clinical, socioeconomic, and behavioural factors at age 50 years and
  risk of cardiometabolic multimorbidity and mortality: {A} cohort study.
\newblock {\em PLoS Medicine}, 15(5):1--16.

\bibitem[{Stan Development Team}, 2022]{Rstan}
{Stan Development Team} (2022).
\newblock {RStan}: {T}he {R} interface to {Stan}.
\newblock R package version 2.21.3.

\bibitem[Vamos et~al., 2012]{vamos2012association}
Vamos, E.~P., Harris, M., Millett, C., Pape, U.~J., Khunti, K., Curcin, V.,
  Molokhia, M., and Majeed, A. (2012).
\newblock Association of systolic and diastolic blood pressure and all cause
  mortality in people with newly diagnosed type 2 diabetes: {R}etrospective
  cohort study.
\newblock {\em BMJ}, 345:1--10.

\bibitem[Vehtari et~al., 2022]{loo}
Vehtari, A., Gabry, J., Magnusson, M., Yao, Y., B{\"u}rkner, P.~C., T., P., and
  Gelman, A. (2022).
\newblock loo: {E}fficient leave-one-out cross-validation and {WAIC} for
  {B}ayesian models.
\newblock R package version 2.5.0.

\bibitem[Vehtari et~al., 2017]{vehtari2017practical}
Vehtari, A., Gelman, A., and Gabry, J. (2017).
\newblock Practical {B}ayesian model evaluation using leave-one-out
  cross-validation and {WAIC}.
\newblock {\em Statistics and Computing}, 27:1413--1432.

\bibitem[Verbeke and Davidian, 2009]{verbeke2009joint}
Verbeke, G. and Davidian, M. (2009).
\newblock {\em Joint models for longitudinal data: {I}ntroduction and
  overview}.
\newblock Chapman \& Hall/CRC.

\bibitem[Wallace et~al., 2015]{wallace2015managing}
Wallace, E., Salisbury, C., Guthrie, B., Lewis, C., Fahey, T., and Smith, S.~M.
  (2015).
\newblock Managing patients with multimorbidity in primary care.
\newblock {\em BMJ}, 350:1--6.

\bibitem[Wolf et~al., 2019]{wolf2019data}
Wolf, A., Dedman, D., Campbell, J., Booth, H., Lunn, D., Chapman, J., and
  Myles, P. (2019).
\newblock Data resource profile: {C}linical practice research datalink ({CPRD})
  aurum.
\newblock {\em International Journal of Epidemiology}, 48(6):1740--1740g.

\end{thebibliography}

\newpage
\appendix
\section{Asymptotic and finite sample comparisons of the inference approaches}
Here, we provide a more detailed comparison between the JM-MSM, JM-CR, and JM-ST approaches by examining the full conditional distributions associated with the working posteriors for each approach. These full conditionals are useful for analyzing the corresponding joint posterior density, as the latter is fully characterized by the former according to the Hammersley-Clifford theorem \citep{robert1999monte}.  For ease of notation and comparison, we assume that, for all approaches, the longitudinal submodel is correctly specified. For the two blockwise approaches, all longitudinal data of subjects associated with a given block would be used. Further, we restrict to MSMs with irreversible transitions so that each subject can only experience a specific type of transition at most once, though the discussion below can carry over to the more general scenario.

Continuing with the notation in Section \ref{sec:jm-msm} of the main paper, let $\Omega=\{(i,j)\mid i\neq j\in S;\ i\to j \ \text{is permitted}\}$. For each type of transition $j\to k$, with $(j,k)\in \Omega$, we introduce a new variable for each subject $i$ as follows. If subject $i$ does not enter state j during the follow-up, we set $D^{(i)}_{jk}=0$. Otherwise, there will exist an $m$ such that $E_{i}(T^{(i)}_{m})=j$, and we define $D^{(i)}_{jk}=T^{(i)}_{m+1}-T^{(i)}_{m}$. Let $\delta^{(i)}_{jk}$ be the indicator function associated with $D^{(i)}_{jk}$, where $\delta^{(i)}_{jk}=1$ if transition $j\to k$ is observed for subject $i$ and is zero if censored (with respect to transition $j\to k$).
For a generic competing risk block $B_{v}$, and a specific transition within the block $j_v \to k$, where $j_v$ is the initial state in the block and $k\in S_{B_v}$, the working posterior associated with the JM-ST approach is proportional to (see Equations \eqref{eq:likUS} and \eqref{eq:likJUS})
\begin{equation}
    \prod_{i \in I_{B_{v}}}f(D^{(i)}_{j_{v}k},\delta^{(i)}_{j_{v}k} \mid \theta^{E}_{j_{v}k},b_{i},\theta_{y})\prod_{i\in I_{B_{v}}}\prod_{j=1}^{n_{i}}f(y_{ij} \mid b_{i},\theta_{y})f(b_{i} \mid \theta_{b})f(\Theta_{B_{j_v,k}}),
    \label{eq:app_JM-ST}
\end{equation}
where \begin{equation*}
f(D^{(i)}_{jk},\delta^{(i)}_{jk} \mid \theta^{E}_{jk},b_{i},\theta_{y})=h^{(i)}_{jk}(D^{(i)}_{jk} \mid \theta^{E}_{jk},b_{i},\theta_{y})^{\delta^{(i)}_{jk}}\exp(-\int_{0}^{D^{(i)}_{jk}}h^{(i)}_{jk}(u \mid \theta^{E}_{jk},b_{i},\theta_{y})du),
\end{equation*} $\theta^{E}_{jk}$ represents transition-specific MSM parameters, $\theta_{y}$ is the parameter vector associated with the longitudinal submodel, $\theta_{b}$ represents parameters associated with the prior distribution of the random effect, and $\Theta_{B_{j_v,k}}=(\theta^{E}_{jk},\theta_{y},\theta_{b})$. The full conditional distribution for each block of parameters, in the log-scale, is given by 
\begin{gather}
\begin{aligned}
   &\log f(\theta^{E}_{j_{v}k} \mid \cdot) = \sum_{i \in I_{B_{v}}}\log f(D^{(i)}_{j_{v}k},\delta^{(i)}_{j_{v}k} \mid \theta^{E}_{j_{v}k},b_{i},\theta_{y})+\log f(\theta^{E}_{j_{v}k})+C,\\
    &\log f(b_{i} \mid \cdot) = \log f(D^{(i)}_{j_{v}k},\delta^{(i)}_{j_{v}k} \mid \theta^{E}_{j_{v}k},b_{i},\theta_{y})+\sum_{j=1}^{n_{i}}\log f(y_{ij} \mid b_{i},\theta_{y})+\log f(b_{i} \mid \theta_{b})+C, \quad i\in I_{B_{v}},\\
    &\log f(\theta_{y} \mid \cdot) = \sum_{i\in I_{B_{v}}}\log f(D^{(i)}_{j_{v}k},\delta^{(i)}_{j_{v}k} \mid \theta^{E}_{j_{v}k},b_{i},\theta_{y})+\sum_{i\in I_{B_{v}}}\sum_{j=1}^{n_{i}}\log f(y_{ij} \mid b_{i},\theta_{y})+\log f(\theta_{y})+C,\\
    &\log f(\theta_{b} \mid \cdot) = \sum_{i\in I_{B_{v}}}\log f(b_{i} \mid \theta_{b})+\log f(\theta_{b})+C,
\end{aligned}
\label{eq:app_JM-ST_fullcond}
\end{gather}
where $C$ denotes a generic constant term (relative to the parameters of interest). The working posterior associated with the JM-CR approach for the block $B_{v}$ is proportional to (see Equations \eqref{eq:likJCR} and \eqref{eq:likCR})
\begin{equation}
    \prod_{(j_{v},k): k \in S_{B_v}}\prod_{i\in I_{B_{v}}}f(D^{(i)}_{j_{v}k},\delta^{(i)}_{j_{v}k} \mid \theta^{E}_{j_{v}k},b_{i},\theta_{y})\prod_{i\in I_{B_{v}}}\prod_{j=1}^{n_{i}}f(y_{ij} \mid b_{i},\theta_{y})f(b_{i} \mid \theta_{b})f(\Theta_{B_{v}}),
     \label{eq:app_JM-CR}
\end{equation}
where $\Theta_{B_{v}}=(\{\theta^{E}_{jk}\}_{k\in S_{B_{v}}},\theta_{y},\theta_{b})$. The associated full conditional distributions are given by
\begin{gather}
\begin{aligned}
    &\log f(\theta^{E}_{j_{v}k} \mid \cdot) = \sum_{i \in I_{B_{v}}}\log f(D^{(i)}_{j_{v}k},\delta^{(i)}_{j_{v}k} \mid \theta^{E}_{j_{v}k},b_{i},\theta_{y})+\log f(\theta^{E}_{j_{v}k})+C, \quad k \in S_{B_v},\\
    &\log f(b_{i} \mid \cdot) = \sum_{(j_{v},k): k \in S_{B_v}}\log f(D^{(i)}_{j_{v}k},\delta^{(i)}_{j_{v}k} \mid \theta^{E}_{j_{v}k},b_{i},\theta_{y})+\sum_{j=1}^{n_{i}}\log f(y_{ij} \mid b_{i},\theta_{y})\\
    &+\log f(b_{i} \mid \theta_{b})+C, \quad i\in I_{B_{v}},\\
    &\log f(\theta_{y} \mid \cdot) = \sum_{i\in I_{B_{v}}}\sum_{(j_{v},k): k \in S_{B_v}}\log f(D^{(i)}_{j_{v}k},\delta^{(i)}_{j_{v}k} \mid \theta^{E}_{j_{v}k},b_{i},\theta_{y})+\sum_{i\in I_{B_{v}}}\sum_{j=1}^{n_{i}}\log f(y_{ij} \mid b_{i},\theta_{y})\\
    &+\log f(\theta_{y})+C,\\
    &\log f(\theta_{b} \mid \cdot) = \sum_{i\in I_{B_{v}}}\log f(b_{i} \mid \theta_{b})+\log f(\theta_{b})+C.
\end{aligned}
\label{eq:app_JM-CR_fullcond}
\end{gather}

Using the newly introduced variables, the working posterior associated with the JM-MSM approach (see Equations \eqref{eq:Post_JMSM} and \eqref{eq:likMSM}), up to a normalizing constant, can be equivalently expressed as
\begin{equation}
\prod_{(j,k)\in \Omega}\prod_{i=1}^{n}f(D^{(i)}_{jk},\delta^{(i)}_{jk} \mid \theta^{E}_{jk},b_{i},\theta_{y})\prod_{i=1}^{n}\prod_{j=1}^{n_{i}}f(y_{ij} \mid b_{i},\theta_{y})f(b_{i} \mid \theta_{b})f(\Theta),
 \label{eq:app_JM-MSM}
\end{equation}
where $\Theta=(\{\theta^{E}_{jk}\}_{(j,k)\in \Omega},\theta_{y},\theta_{b})$. The associated full conditional distributions are given by
\begin{gather}
\begin{aligned}
   & \log f(\theta^{E}_{j_{v}k} \mid \cdot) = \sum_{i \in I_{B_{v}}}\log f(D^{(i)}_{j_{v}k},\delta^{(i)}_{j_{v}k} \mid \theta^{E}_{j_{v}k},b_{i},\theta_{y})+\log f(\theta^{E}_{j_{v}k})+C,\\
   & \log f(b_{i} \mid \cdot) = \sum_{(j,k)\in \Omega}\log f(D^{(i)}_{jk},\delta^{(i)}_{jk} \mid \theta^{E}_{jk},b_{i},\theta_{y})+\sum_{j=1}^{n_{i}}\log f(y_{ij} \mid b_{i},\theta_{y})+\log f(b_{i} \mid \theta_{b})+C, \\
   & \log f(\theta_{y} \mid \cdot) = \sum_{i=1}^{n}\sum_{(j,k)\in \Omega}\log f(D^{(i)}_{jk},\delta^{(i)}_{jk} \mid \theta^{E}_{jk},b_{i},\theta_{y})+\sum_{i=1}^{n}\sum_{j=1}^{n_{i}}\log f(y_{ij} \mid b_{i},\theta_{y})+\log f(\theta_{y})+C,\\
   & \log f(\theta_{b} \mid \cdot) = \sum_{i=1}^{n}\log f(b_{i} \mid \theta_{b})+\log f(\theta_{b})+C.
\end{aligned}
\label{eq:app_JM-MSM_fullcond}
\end{gather}

Full conditionals for MSM parameters associated with other types of transitions exhibit a similar form and are therefore omitted here.

We now analyze the behaviours of the marginal posteriors for the transition-specific MSM parameters, $\theta^{E}_{jk}$, associated with the posteriors as defined in Equations \eqref{eq:app_JM-ST}, \eqref{eq:app_JM-CR}, and \eqref{eq:app_JM-MSM}, as $n$ and $n_{i}$ grow. We proceed by examining the behaviours of the full conditionals associated with \eqref{eq:app_JM-ST}, \eqref{eq:app_JM-CR}, and \eqref{eq:app_JM-MSM}. 
First, note that the full conditional distribution for $\theta^{E}_{jk}$ is exactly the same across three approaches (see the first line of Equations \eqref{eq:app_JM-ST_fullcond}, \eqref{eq:app_JM-CR_fullcond}, and \eqref{eq:app_JM-MSM_fullcond}). With MSM structure fixed (and thus the cardinality of $\Omega$), we observe that as $n_{i}$ increases, inference from JM-ST and JM-CR would increasingly resemble each other (compare Equations \eqref{eq:app_JM-ST_fullcond} and \eqref{eq:app_JM-CR_fullcond}), as the leading term in $\log f(b_{i} \mid \cdot)$ and $\log f(\theta_{y} \mid \cdot)$ is the log-density of the longitudinal submodel, and the full conditional for $\theta_{b}$ is exactly the same under the two approaches. For JM-MSM in \eqref{eq:app_JM-MSM_fullcond}, the contribution from the longitudinal submodel would also dominate as $n_{i}$ increases.
If in addition, we let $n$ increase, so that the cardinality of $I_{B_{v}}$ also increases, then under some regularity conditions for the densities of the longitudinal submodel and random effects (which hold for densities in the exponential family), the full conditionals associated with the three approaches would have the same asymptotic distribution according to Bayesian asymptotic theory (see e.g. Chapter 4 of \citet{gelman2013bayesian} and Chapter 10 of \citet{cox1979theoretical}). Therefore, the marginal posterior inference for $\theta^{E}_{j_{v}k}$ (and indeed also for $\theta_{y}$ and $\theta_{b}$) would be asymptotically equivalent under the JM-MSM, JM-CR, and JM-ST approaches.
For a given dataset (i.e. fixed $n$ and $n_{i}$), the blockwise approaches to posterior inference for a specific block/transition essentially perform inference based solely on the 'relevant' subset of the entire dataset, specifically the data from subjects who are at risk of a transition in the block.
As a result, the posterior variability of parameters obtained using blockwise approaches would be expected to be larger than using the JM-MSM approach, which utilizes all data. However, we show in the simulation study that, with a moderate dataset, the blockwise approaches provide good point and interval estimation properties as compared to the JM-MSM approach.

\newpage

\section{Additional results for the simulation study}
Here we provide additional figures and tables showing the estimation results of the MSM parameters for each simulation model, complementing Section \ref{subsec:sim_results} of the main paper.

\begin{figure}[htbp]
  \centering
\includegraphics[width=1\textwidth]{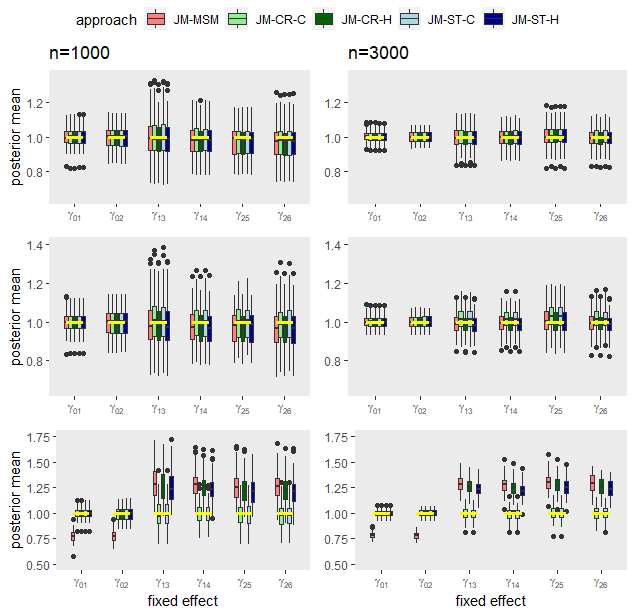}
 \caption{Box plot summary of the posterior mean of the fixed effect $\gamma_{jk}$ obtained by each approach from $100$ replications of the data simulated from Model 1. The settings are the same as in Figure \ref{fig:sim1results}.}
\end{figure}

\begin{figure}[htbp]
  \centering
\includegraphics[width=1\textwidth]{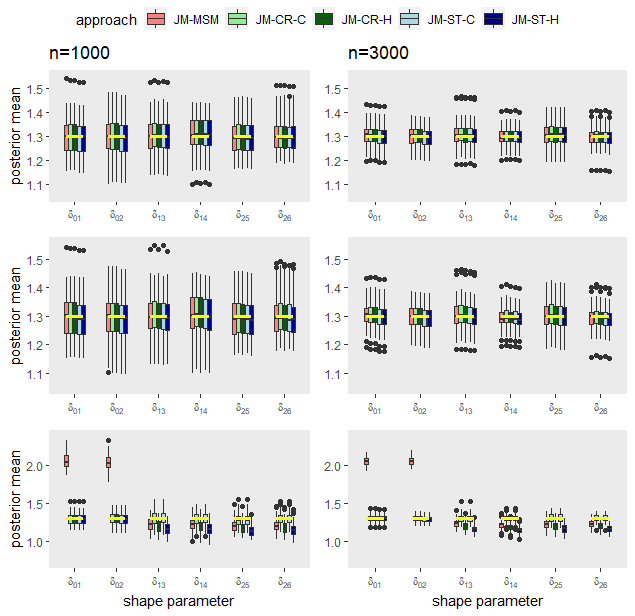}
 \caption{Box plot summary of the posterior mean of the Weibull shape parameter $\delta_{jk}$ obtained by each approach from $100$ replications of the data simulated from Model 1. The settings are the same as in Figure \ref{fig:sim1results}.}
\end{figure}

\begin{figure}[htbp]
  \centering
\includegraphics[width=1\textwidth]{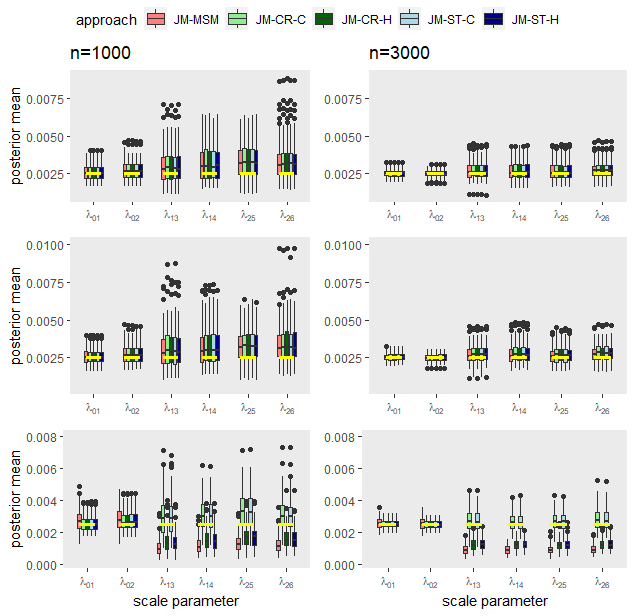}
 \caption{Box plot summary of the posterior mean of the Weibull scale parameter$\lambda_{jk}$ obtained by each approach from $100$ replications of the data simulated from Model 1. The settings are the same as in Figure \ref{fig:sim1results}.}
\end{figure}

\begin{figure}[htbp]
  \centering
\includegraphics[width=1\textwidth]{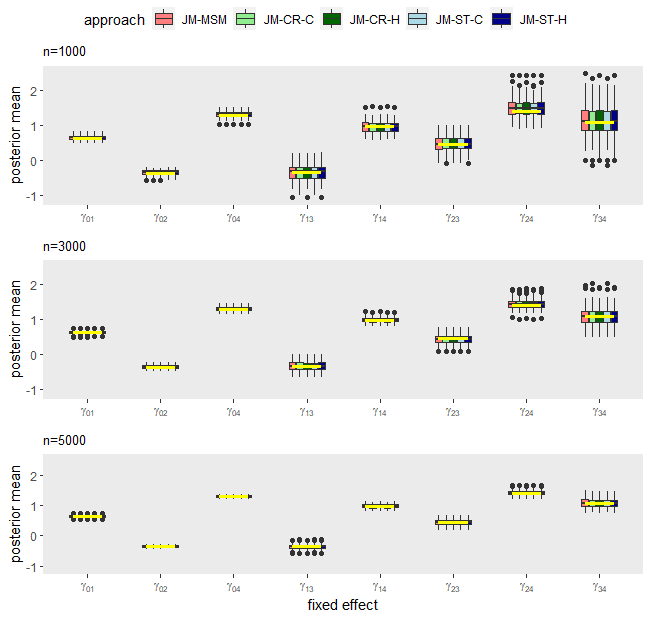}
 \caption{Box plot summary of the posterior mean of the fixed effect $\gamma_{jk}$ obtained by each approach from $100$ replications of the data simulated from Model 2. The settings are the same as in Figure \ref{fig:sim2results}.}
\end{figure}

\begin{figure}[htbp]
  \centering
\includegraphics[width=1\textwidth]{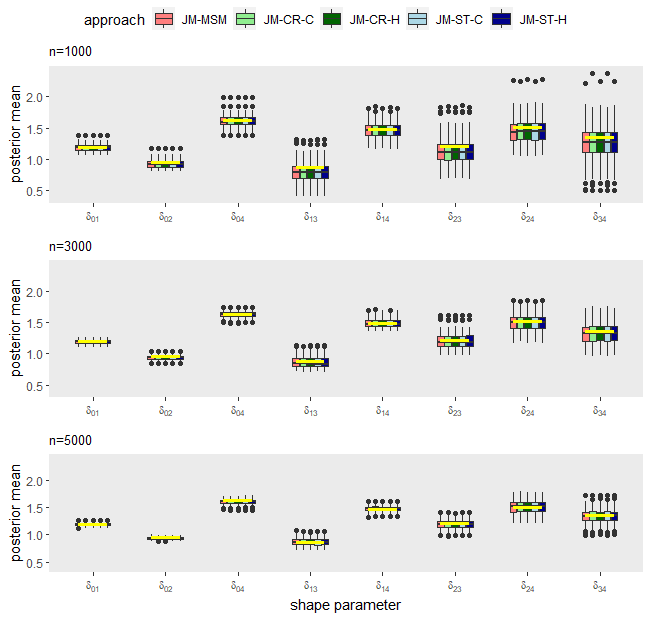}
 \caption{Box plot summary of the posterior mean of the Weibull shape parameter $\delta_{jk}$ obtained by each approach from $100$ replications of the data simulated from Model 2. The settings are the same as in Figure \ref{fig:sim2results}.}
\end{figure}

\begin{figure}[htbp]
  \centering
\includegraphics[width=1\textwidth]{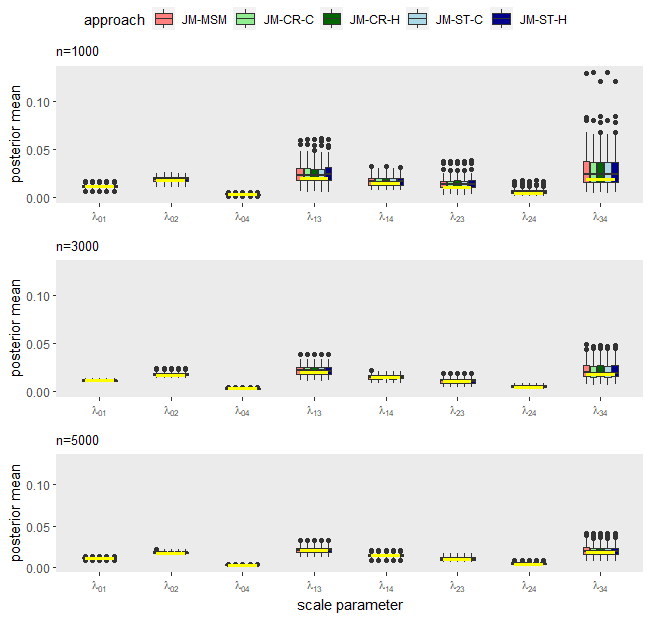}
 \caption{Box plot summary of the posterior mean of the Weibull scale parameter $\lambda_{jk}$ obtained by each approach from $100$ replications of the data simulated from Model 2. The settings are the same as in Figure \ref{fig:sim2results}.}
\end{figure}

\begin{table}[htbp]
\centering
\caption{Coverage probability of estimated $95\%$ credible interval for the fixed effect based on $100$ replications of data from simulation Model 1 ($n=1000$ \textbar{} $n=3000$).}
\begin{tabular}{@{}ccccc@{}}
\toprule
Parameter & \multicolumn{1}{c}{Approach} & \multicolumn{1}{c}{Scenario 1} & \multicolumn{1}{c}{Scenario 2} & \multicolumn{1}{c}{Scenario 3} \\ \midrule
\multirow{5}{*}{$\gamma_{01}$} & JM-MSM  & 0.96 \textbar{} 0.96  & 0.97 \textbar{} 0.96  & 0.02 \textbar{} 0.00 \\
                            & JM-CR-C & 0.96 \textbar{} 0.96 & 0.98 \textbar{} 0.96 & 0.97 \textbar{} 0.96\\
                            & JM-CR-H & 0.96 \textbar{} 0.96 & 0.98 \textbar{} 0.96 & 0.97 \textbar{} 0.96\\
                            & JM-ST-C & 0.97 \textbar{} 0.96 & 0.97 \textbar{} 0.96& 0.98 \textbar{} 0.95\\
                            & JM-ST-H & 0.97 \textbar{} 0.96 & 0.97 \textbar{} 0.96& 0.98 \textbar{} 0.95\\\addlinespace
\multirow{5}{*}{$\gamma_{02}$} & JM-MSM  & 0.92 \textbar{} 0.99 & 0.92 \textbar{} 0.99& 0.03 \textbar{} 0.00\\
                            & JM-CR-C & 0.93 \textbar{} 0.98 & 0.93 \textbar{} 0.98& 0.93 \textbar{} 0.99\\
                            & JM-CR-H & 0.93 \textbar{} 0.98 & 0.93 \textbar{} 0.98& 0.93 \textbar{} 0.99\\
                            & JM-ST-C & 0.95 \textbar{} 1.00 & 0.92 \textbar{} 1.00& 0.90 \textbar{} 0.99\\
                            & JM-ST-H & 0.95 \textbar{} 1.00 & 0.92 \textbar{} 1.00&  0.90 \textbar{} 0.99\\\addlinespace
\multirow{5}{*}{$\gamma_{13}$} & JM-MSM  & 0.92 \textbar{} 0.96 & 0.91 \textbar{} 0.95& 0.44 \textbar{} 0.04\\
                            & JM-CR-C & 0.92 \textbar{} 0.96 & 0.90 \textbar{} 0.94& 0.90 \textbar{} 0.96\\
                            & JM-CR-H & 0.92 \textbar{} 0.95  & 0.92 \textbar{} 0.96& 0.53 \textbar{} 0.07\\
                            & JM-ST-C & 0.92 \textbar{} 0.96& 0.91 \textbar{} 0.97& 0.93 \textbar{} 0.97\\
                            & JM-ST-H & 0.92 \textbar{} 0.96& 0.90 \textbar{} 0.97& 0.56 \textbar{} 0.13\\\addlinespace
\multirow{5}{*}{$\gamma_{14}$} & JM-MSM  & 0.96 \textbar{} 0.95& 0.96 \textbar{} 0.95& 0.42 \textbar{} 0.06\\
                            & JM-CR-C & 0.98 \textbar{} 0.98& 0.98 \textbar{} 0.98& 0.96 \textbar{} 0.92\\
                            & JM-CR-H & 0.96 \textbar{} 0.95& 0.96 \textbar{} 0.94& 0.58 \textbar{} 0.11\\
                            & JM-ST-C & 0.96 \textbar{} 0.96& 0.97 \textbar{} 0.98& 0.97 \textbar{} 0.94\\
                            & JM-ST-H & 0.95 \textbar{} 0.97& 0.96 \textbar{} 0.97& 0.63 \textbar{} 0.14\\\addlinespace
\multirow{5}{*}{$\gamma_{25}$} & JM-MSM  & 0.97 \textbar{} 0.95 & 0.99 \textbar{} 0.95 & 0.51 \textbar{} 0.02\\
                            & JM-CR-C & 0.99 \textbar{} 0.94 & 1.00 \textbar{} 0.89 & 0.95 \textbar{} 0.92\\
                            & JM-CR-H & 0.97 \textbar{} 0.96 & 0.97 \textbar{} 0.96 & 0.58 \textbar{} 0.05\\
                            & JM-ST-C & 0.97 \textbar{} 0.92 & 1.00 \textbar{} 0.89 & 0.93 \textbar{} 0.93\\
                            & JM-ST-H & 0.97 \textbar{} 0.96 & 0.98 \textbar{} 0.95 & 0.65 \textbar{} 0.08\\\addlinespace
\multirow{5}{*}{$\gamma_{26}$} & JM-MSM  & 0.94 \textbar{} 0.95 & 0.94 \textbar{} 0.94 & 0.55 \textbar{} 0.04\\
                            & JM-CR-C & 0.95 \textbar{} 0.94 & 0.92 \textbar{} 0.96 & 0.94 \textbar{} 0.94\\
                            & JM-CR-H & 0.95 \textbar{} 0.94 & 0.96 \textbar{} 0.93 & 0.63 \textbar{} 0.06\\
                            & JM-ST-C & 0.94 \textbar{} 0.94 & 0.93 \textbar{} 0.95 & 0.94 \textbar{} 0.94\\
                            & JM-ST-H & 0.94 \textbar{} 0.93 & 0.93 \textbar{} 0.94 & 0.69 \textbar{} 0.17\\\bottomrule
\end{tabular}
\end{table}

\begin{table}[htbp]
\centering
\caption{Coverage probability of estimated $95\%$ credible interval for the shape parameter based on $100$ replications of data from simulation Model 1 ($n=1000$ \textbar{} $n=3000$).}
\begin{tabular}{@{}ccccc@{}}
\toprule
Parameter & \multicolumn{1}{c}{Approach} & \multicolumn{1}{c}{Scenario 1} & \multicolumn{1}{c}{Scenario 2} & \multicolumn{1}{c}{Scenario 3} \\ \midrule
\multirow{5}{*}{$\delta_{01}$} & JM-MSM  & 0.97 \textbar{} 0.96  & 0.97 \textbar{} 0.96  & 0.00 \textbar{} 0.00 \\
                            & JM-CR-C & 0.98 \textbar{} 0.96 & 0.96 \textbar{} 0.95 & 0.96 \textbar{} 0.96\\
                            & JM-CR-H & 0.98 \textbar{} 0.96 & 0.96 \textbar{} 0.95 & 0.96 \textbar{} 0.96\\
                            & JM-ST-C & 0.96 \textbar{} 0.95 & 0.97 \textbar{} 0.94& 0.96 \textbar{} 0.95\\
                            & JM-ST-H & 0.96 \textbar{} 0.95 & 0.97 \textbar{} 0.94& 0.96 \textbar{} 0.95\\\addlinespace
\multirow{5}{*}{$\delta_{02}$} & JM-MSM  & 0.91 \textbar{} 0.93 & 0.92 \textbar{} 0.95& 0.00 \textbar{} 0.00\\
                            & JM-CR-C & 0.91 \textbar{} 0.95 & 0.92 \textbar{} 0.94& 0.93 \textbar{} 0.93\\
                            & JM-CR-H & 0.91 \textbar{} 0.95 & 0.92 \textbar{} 0.94& 0.93 \textbar{} 0.93\\
                            & JM-ST-C & 0.92 \textbar{} 0.97 & 0.90 \textbar{} 0.95& 0.91 \textbar{} 0.93\\
                            & JM-ST-H & 0.92 \textbar{} 0.97 & 0.90 \textbar{} 0.95&  0.91 \textbar{} 0.93\\\addlinespace
\multirow{5}{*}{$\delta_{13}$} & JM-MSM  & 0.93 \textbar{} 0.95 & 0.94 \textbar{} 0.93 & 0.81 \textbar{} 0.68 \\
                            & JM-CR-C & 0.94 \textbar{} 0.94 & 0.93 \textbar{} 0.93 & 0.92 \textbar{} 0.92 \\
                            & JM-CR-H & 0.95 \textbar{} 0.95  & 0.93 \textbar{} 0.95 & 0.73 \textbar{} 0.53 \\
                            & JM-ST-C & 0.93 \textbar{} 0.92 & 0.94 \textbar{} 0.94 & 0.90 \textbar{} 0.94 \\
                            & JM-ST-H & 0.96 \textbar{} 0.95 & 0.93 \textbar{} 0.94 & 0.58 \textbar{} 0.19 \\\addlinespace
\multirow{5}{*}{$\delta_{14}$} & JM-MSM  & 0.96 \textbar{} 0.97 & 0.97 \textbar{} 0.97 & 0.88 \textbar{} 0.63 \\
                            & JM-CR-C & 0.96 \textbar{} 0.96 & 0.96 \textbar{} 0.97 & 0.95 \textbar{} 0.96 \\
                            & JM-CR-H & 0.96 \textbar{} 0.97 & 0.95 \textbar{} 0.96 & 0.78 \textbar{} 0.38 \\
                            & JM-ST-C & 0.96 \textbar{} 0.96 & 0.95 \textbar{} 0.95 & 0.97 \textbar{} 0.96 \\
                            & JM-ST-H & 0.95 \textbar{} 0.96 & 0.96 \textbar{} 0.97 & 0.61 \textbar{} 0.14 \\\addlinespace
\multirow{5}{*}{$\delta_{25}$} & JM-MSM  & 0.96 \textbar{} 0.91 & 0.98 \textbar{} 0.90 & 0.77 \textbar{} 0.66 \\
                            & JM-CR-C & 0.97 \textbar{} 0.90 & 0.96 \textbar{} 0.88 & 0.97 \textbar{} 0.93 \\
                            & JM-CR-H & 0.99 \textbar{} 0.89 & 0.96 \textbar{} 0.91 & 0.70 \textbar{} 0.52 \\
                            & JM-ST-C & 0.97 \textbar{} 0.90 & 0.98 \textbar{} 0.89 & 0.96 \textbar{} 0.93 \\
                            & JM-ST-H & 0.97 \textbar{} 0.90 & 0.97 \textbar{} 0.90 & 0.51 \textbar{} 0.27 \\\addlinespace
\multirow{5}{*}{$\delta_{26}$} & JM-MSM  & 0.97 \textbar{} 0.97 & 0.95 \textbar{} 0.94 & 0.81 \textbar{} 0.70 \\
                            & JM-CR-C & 0.96 \textbar{} 0.98 & 0.93 \textbar{} 0.94 & 0.93 \textbar{} 0.98 \\
                            & JM-CR-H & 0.97 \textbar{} 0.98 & 0.96 \textbar{} 0.97 & 0.70 \textbar{} 0.48 \\
                            & JM-ST-C & 0.96 \textbar{} 0.97 & 0.94 \textbar{} 0.96 & 0.93 \textbar{} 0.97 \\
                            & JM-ST-H & 0.96 \textbar{} 0.97 & 0.96 \textbar{} 0.95 & 0.55 \textbar{} 0.16 \\\bottomrule
\end{tabular}
\end{table}

\begin{table}[htbp]
\centering
\caption{Coverage probability of estimated $95\%$ credible interval for the scale parameter based on $100$ replications of data from simulation Model 1 ($n=1000$ \textbar{} $n=3000$).}
\begin{tabular}{@{}ccccc@{}}
\toprule
Parameter & \multicolumn{1}{c}{Approach} & \multicolumn{1}{c}{Scenario 1} & \multicolumn{1}{c}{Scenario 2} & \multicolumn{1}{c}{Scenario 3} \\ \midrule
\multirow{5}{*}{$\lambda_{01}$} & JM-MSM  & 0.97 \textbar{} 0.96  & 0.96 \textbar{} 0.98  & 0.89 \textbar{} 0.97 \\
                            & JM-CR-C & 0.97 \textbar{} 0.98 & 0.96 \textbar{} 0.96 & 0.96 \textbar{} 0.97\\
                            & JM-CR-H & 0.97 \textbar{} 0.98 & 0.96 \textbar{} 0.96 & 0.96 \textbar{} 0.97\\
                            & JM-ST-C & 0.97 \textbar{} 0.98 & 0.97 \textbar{} 0.97& 0.97 \textbar{} 0.97\\
                            & JM-ST-H & 0.97 \textbar{} 0.98 & 0.97 \textbar{} 0.97& 0.97 \textbar{} 0.97\\\addlinespace
\multirow{5}{*}{$\lambda_{02}$} & JM-MSM  & 0.92 \textbar{} 0.98 & 0.93 \textbar{} 0.98& 0.90 \textbar{} 0.94\\
                            & JM-CR-C & 0.93 \textbar{} 0.97 & 0.93 \textbar{} 0.98& 0.92 \textbar{} 0.97\\
                            & JM-CR-H & 0.93 \textbar{} 0.97 & 0.93 \textbar{} 0.98& 0.92 \textbar{} 0.97\\
                            & JM-ST-C & 0.92 \textbar{} 0.97 & 0.94 \textbar{} 0.99& 0.93 \textbar{} 0.96\\
                            & JM-ST-H & 0.92 \textbar{} 0.97 & 0.94 \textbar{} 0.99&  0.93 \textbar{} 0.96\\\addlinespace
\multirow{5}{*}{$\lambda_{13}$} & JM-MSM  & 0.95 \textbar{} 0.94 & 0.95 \textbar{} 0.95 & 0.42 \textbar{} 0.03\\
                            & JM-CR-C & 0.96 \textbar{} 0.95 & 0.94 \textbar{} 0.96 & 0.91 \textbar{} 0.92\\
                            & JM-CR-H & 0.94 \textbar{} 0.93 & 0.95 \textbar{} 0.95 & 0.68 \textbar{} 0.19\\
                            & JM-ST-C & 0.93 \textbar{} 0.96 & 0.93 \textbar{} 0.95 & 0.91 \textbar{} 0.92\\
                            & JM-ST-H & 0.94 \textbar{} 0.94 & 0.94 \textbar{} 0.96 & 0.67 \textbar{} 0.23\\\addlinespace
\multirow{5}{*}{$\lambda_{14}$} & JM-MSM  & 0.95 \textbar{} 0.95 & 0.93 \textbar{} 0.96 & 0.51 \textbar{} 0.03\\
                            & JM-CR-C & 0.94 \textbar{} 0.94 & 0.92 \textbar{} 0.96 & 0.98 \textbar{} 0.95\\
                            & JM-CR-H & 0.95 \textbar{} 0.96 & 0.91 \textbar{} 0.95 & 0.74 \textbar{} 0.17\\
                            & JM-ST-C & 0.93 \textbar{} 0.94 & 0.92 \textbar{} 0.96 & 0.98 \textbar{} 0.94\\
                            & JM-ST-H & 0.94 \textbar{} 0.94 & 0.91 \textbar{} 0.94 & 0.71 \textbar{} 0.14\\\addlinespace
\multirow{5}{*}{$\lambda_{25}$} & JM-MSM  & 0.97 \textbar{} 0.94 & 0.95 \textbar{} 0.91 & 0.62 \textbar{} 0.03\\
                            & JM-CR-C & 0.95 \textbar{} 0.92 & 0.96 \textbar{} 0.93 & 0.96 \textbar{} 0.91\\
                            & JM-CR-H & 0.95 \textbar{} 0.93 & 0.93 \textbar{} 0.92 & 0.82 \textbar{} 0.15\\
                            & JM-ST-C & 0.96 \textbar{} 0.93 & 0.97 \textbar{} 0.93 & 0.94 \textbar{} 0.94\\
                            & JM-ST-H & 0.94 \textbar{} 0.93 & 0.94 \textbar{} 0.92 & 0.82 \textbar{} 0.16\\\addlinespace
\multirow{5}{*}{$\lambda_{26}$} & JM-MSM  & 0.93 \textbar{} 0.95 & 0.96 \textbar{} 0.95 & 0.53 \textbar{} 0.04\\
                            & JM-CR-C & 0.92 \textbar{} 0.94 & 0.94 \textbar{} 0.93 & 0.97 \textbar{} 0.93\\
                            & JM-CR-H & 0.93 \textbar{} 0.93 & 0.94 \textbar{} 0.94 & 0.78 \textbar{} 0.20\\
                            & JM-ST-C & 0.94 \textbar{} 0.94 & 0.94 \textbar{} 0.95 & 0.95 \textbar{} 0.95\\
                            & JM-ST-H & 0.94 \textbar{} 0.96 & 0.95 \textbar{} 0.94 & 0.76 \textbar{} 0.24\\\bottomrule
\end{tabular}
\end{table}


\begin{table}[htbp]
\centering
\caption{Coverage probability of estimated $95\%$ credible interval for the fixed effect based on $100$ replications of data from simulation Model 2 (Note: results for MSM with $n=5000$ are based on the $99$ available repetitions).}
\begin{tabular}{@{}ccccc@{}}
\toprule
Parameter & \multicolumn{1}{c}{Approach} & \multicolumn{1}{c}{$n=1000$} & \multicolumn{1}{c}{$n=3000$} & \multicolumn{1}{c}{$n=5000$} \\ \midrule
\multirow{5}{*}{$\gamma_{01}$} & JM-MSM  & 0.98  & 0.90  & 0.93  \\
                              & JM-CR-C & 0.99 & 0.90 & 0.95 \\
                              & JM-CR-H & 0.99 & 0.90 & 0.95 \\
                              & JM-ST-C & 0.98 & 0.90 & 0.94 \\
                              & JM-ST-H & 0.98 & 0.90 & 0.94 \\\addlinespace
\multirow{5}{*}{$\gamma_{02}$} & JM-MSM  & 0.95 & 0.93 & 0.96 \\
                              & JM-CR-C & 0.95 & 0.92 & 0.95 \\
                              & JM-CR-H & 0.95 & 0.92 & 0.95 \\
                              & JM-ST-C & 0.95 & 0.93 & 0.97 \\
                              & JM-ST-H & 0.95 & 0.93 & 0.97 \\\addlinespace
\multirow{5}{*}{$\gamma_{04}$} & JM-MSM  & 0.95  & 0.94  & 0.98  \\
                             & JM-CR-C & 0.96 & 0.93 & 0.98 \\
                             & JM-CR-H & 0.96 & 0.93 & 0.98 \\
                             & JM-ST-C & 0.94 & 0.94 & 0.98 \\
                             & JM-ST-H & 0.94 & 0.94 & 0.98 \\\addlinespace
\multirow{5}{*}{$\gamma_{13}$} & JM-MSM  & 0.97 & 0.92 & 0.95 \\
                             & JM-CR-C & 0.98 & 0.92 & 0.96 \\
                             & JM-CR-H & 0.97 & 0.92 & 0.97 \\
                             & JM-ST-C & 0.97 & 0.92 & 0.97 \\
                             & JM-ST-H & 0.96 & 0.93 & 0.98 \\\addlinespace
\multirow{5}{*}{$\gamma_{14}$} & JM-MSM  & 0.96  & 0.97  & 0.97  \\
                             & JM-CR-C & 0.96 & 0.97 & 0.98 \\
                             & JM-CR-H & 0.95 & 0.96 & 0.97 \\
                             & JM-ST-C & 0.95 & 0.96 & 0.97 \\
                             & JM-ST-H & 0.96 & 0.97 & 0.97 \\\addlinespace
\multirow{5}{*}{$\gamma_{23}$} & JM-MSM  & 0.95 & 0.94 & 0.91 \\
                             & JM-CR-C & 0.93 & 0.93 & 0.92 \\
                             & JM-CR-H & 0.96 & 0.92 & 0.93 \\
                             & JM-ST-C & 0.95 & 0.95 & 0.92 \\
                             & JM-ST-H & 0.95 & 0.93 & 0.93 \\\addlinespace
\multirow{5}{*}{$\gamma_{24}$} & JM-MSM  & 0.94 & 0.94 & 0.96 \\
                             & JM-CR-C & 0.95 & 0.94 & 0.96 \\
                             & JM-CR-H & 0.94 & 0.96 & 0.95 \\
                             & JM-ST-C & 0.95 & 0.95 & 0.96 \\
                             & JM-ST-H & 0.94 & 0.96 & 0.97 \\\addlinespace    
\multirow{5}{*}{$\gamma_{34}$} & JM-MSM  & 0.95 & 0.94 & 0.97 \\
                             & JM-CR-C & 0.96 & 0.93 & 0.96 \\
                             & JM-CR-H & 0.96 & 0.94 & 0.97 \\
                             & JM-ST-C & 0.96 & 0.93 & 0.96 \\
                             & JM-ST-H & 0.96 & 0.94 & 0.97\\\bottomrule
\end{tabular}
\end{table}

\begin{table}[htbp]
\centering
\caption{Coverage probability of estimated $95\%$ credible interval for the shape parameter based on $100$ replications of data from simulation Model 2 (Note: results for MSM with $n=5000$ are based on the $99$ available repetitions).}
\begin{tabular}{@{}ccccc@{}}
\toprule
Parameter & \multicolumn{1}{c}{Approach} & \multicolumn{1}{c}{$n=1000$} & \multicolumn{1}{c}{$n=3000$} & \multicolumn{1}{c}{$n=5000$} \\ \midrule
\multirow{5}{*}{$\delta_{01}$} & JM-MSM  & 0.99  & 0.98  & 0.95  \\
                              & JM-CR-C & 0.99 & 0.99 & 0.95 \\
                              & JM-CR-H & 0.99 & 0.99 & 0.95 \\
                              & JM-ST-C & 0.98 & 0.99 & 0.95 \\
                              & JM-ST-H & 0.98 & 0.99 & 0.95 \\\addlinespace
\multirow{5}{*}{$\delta_{02}$} & JM-MSM  & 0.93 & 0.97 & 0.94 \\
                              & JM-CR-C & 0.92 & 0.95 & 0.95 \\
                              & JM-CR-H & 0.92 & 0.95 & 0.95 \\
                              & JM-ST-C & 0.89 & 0.93 & 0.93 \\
                              & JM-ST-H & 0.89 & 0.93 & 0.93 \\\addlinespace
\multirow{5}{*}{$\delta_{04}$} & JM-MSM  & 0.92 & 0.92 & 0.87 \\
                             & JM-CR-C & 0.92 & 0.93 & 0.89 \\
                             & JM-CR-H & 0.92 & 0.93 & 0.89 \\
                             & JM-ST-C & 0.92 & 0.93 & 0.89 \\
                             & JM-ST-H & 0.92 & 0.93 & 0.89 \\\addlinespace
\multirow{5}{*}{$\delta_{13}$} & JM-MSM  & 0.93 & 0.97 & 0.96 \\
                             & JM-CR-C & 0.93 & 0.97 & 0.95 \\
                             & JM-CR-H & 0.93 & 0.97 & 0.96 \\
                             & JM-ST-C & 0.93 & 0.97 & 0.95 \\
                             & JM-ST-H & 0.93 & 0.97 & 0.96 \\\addlinespace
\multirow{5}{*}{$\delta_{14}$} & JM-MSM  & 0.92 & 0.94 & 0.93 \\
                             & JM-CR-C & 0.92 & 0.95 & 0.93 \\
                             & JM-CR-H & 0.91 & 0.94 & 0.92 \\
                             & JM-ST-C & 0.91 & 0.96 & 0.94 \\
                             & JM-ST-H & 0.92 & 0.94 & 0.94 \\\addlinespace
\multirow{5}{*}{$\delta_{23}$} & JM-MSM  & 0.89 & 0.94 & 0.95 \\
                             & JM-CR-C & 0.87 & 0.93 & 0.96 \\
                             & JM-CR-H & 0.88 & 0.95 & 0.96 \\
                             & JM-ST-C & 0.86 & 0.93 & 0.95 \\
                             & JM-ST-H & 0.90 & 0.94 & 0.96 \\\addlinespace
\multirow{5}{*}{$\delta_{24}$} & JM-MSM  & 0.93 & 0.93 & 0.95 \\
                             & JM-CR-C & 0.94 & 0.93 & 0.95 \\
                             & JM-CR-H & 0.94 & 0.93 & 0.94 \\
                             & JM-ST-C & 0.93 & 0.92 & 0.94 \\
                             & JM-ST-H & 0.95 & 0.94 & 0.94 \\\addlinespace
\multirow{5}{*}{$\delta_{34}$} & JM-MSM  & 0.93 & 0.94 & 0.95 \\
                             & JM-CR-C & 0.93 & 0.93 & 0.95 \\
                             & JM-CR-H & 0.94 & 0.94 & 0.95 \\
                             & JM-ST-C & 0.93 & 0.93 & 0.95 \\
                             & JM-ST-H & 0.94 & 0.94 & 0.95 \\\bottomrule
\end{tabular}
\end{table}

\begin{table}[htbp]
\centering
\caption{Coverage probability of estimated $95\%$ credible interval for the scale parameter based on $100$ replications of data from simulation Model 2 (Note: results for MSM with $n=5000$ are based on the $99$ available repetitions).}
\begin{tabular}{@{}ccccc@{}}
\toprule
Parameter & \multicolumn{1}{c}{Approach} & \multicolumn{1}{c}{$n=1000$} & \multicolumn{1}{c}{$n=3000$} & \multicolumn{1}{c}{$n=5000$} \\ \midrule
\multirow{5}{*}{$\lambda_{01}$} & JM-MSM  & 0.98  & 1.00  & 0.97  \\
                              & JM-CR-C & 0.98 & 1.00 & 0.97 \\
                              & JM-CR-H & 0.98 & 1.00 & 0.97 \\
                              & JM-ST-C & 0.98 & 1.00 & 0.96 \\
                              & JM-ST-H & 0.98 & 1.00 & 0.96 \\\addlinespace
\multirow{5}{*}{$\lambda_{02}$} & JM-MSM  & 0.95 & 0.93 & 0.96 \\
                              & JM-CR-C & 0.94 & 0.93 & 0.96 \\
                              & JM-CR-H & 0.94 & 0.93 & 0.96 \\
                              & JM-ST-C & 0.94 & 0.92 & 0.95 \\
                              & JM-ST-H & 0.94 & 0.92 & 0.95 \\\addlinespace
\multirow{5}{*}{$\lambda_{04}$} & JM-MSM  & 0.92 & 0.94 & 0.87 \\
                             & JM-CR-C & 0.92 & 0.93 & 0.86 \\
                             & JM-CR-H & 0.92 & 0.93 & 0.86 \\
                             & JM-ST-C & 0.92 & 0.94 & 0.88 \\
                             & JM-ST-H & 0.92 & 0.94 & 0.88 \\\addlinespace
\multirow{5}{*}{$\lambda_{13}$} & JM-MSM  & 0.94 & 0.95 & 0.96 \\
                             & JM-CR-C & 0.93 & 0.94 & 0.96 \\
                             & JM-CR-H & 0.94 & 0.96 & 0.95 \\
                             & JM-ST-C & 0.95 & 0.95 & 0.95 \\
                             & JM-ST-H & 0.93 & 0.94 & 0.95 \\\addlinespace
\multirow{5}{*}{$\lambda_{14}$} & JM-MSM  & 0.96 & 0.96 & 0.95 \\
                             & JM-CR-C & 0.93 & 0.97 & 0.94 \\
                             & JM-CR-H & 0.93 & 0.95 & 0.95 \\
                             & JM-ST-C & 0.92 & 0.96 & 0.94 \\
                             & JM-ST-H & 0.95 & 0.96 & 0.96 \\\addlinespace
\multirow{5}{*}{$\lambda_{23}$} & JM-MSM  & 0.95 & 0.94 & 0.97 \\
                             & JM-CR-C & 0.93 & 0.94 & 0.96 \\
                             & JM-CR-H & 0.95 & 0.94 & 0.96\\
                             & JM-ST-C & 0.94 & 0.93 & 0.96 \\
                             & JM-ST-H & 0.95 & 0.94 & 0.97 \\\addlinespace
\multirow{5}{*}{$\lambda_{24}$} & JM-MSM  & 0.94 & 0.95 & 0.95 \\
                             & JM-CR-C & 0.94 & 0.96 & 0.95 \\
                             & JM-CR-H & 0.96 & 0.95 & 0.95 \\
                             & JM-ST-C & 0.96 & 0.96 & 0.96 \\
                             & JM-ST-H & 0.95 & 0.96 & 0.95 \\\addlinespace
\multirow{5}{*}{$\lambda_{34}$} & JM-MSM  & 0.95 & 0.95 & 0.95 \\
                             & JM-CR-C & 0.96 & 0.93 & 0.93 \\
                             & JM-CR-H & 0.95 & 0.93 & 0.93 \\
                             & JM-ST-C & 0.96 & 0.93 & 0.93 \\
                             & JM-ST-H & 0.95 & 0.93 & 0.93 \\\bottomrule
\end{tabular}
\end{table}

\end{document}